\documentclass[useAMS,usenatbib]{mn2e}
\usepackage{graphicx}
\usepackage{txfonts}
\begin{document}

\title[Scattering of ten pulsars ]
{The study of multi-frequency scattering of ten radio pulsars}
\author[Wojciech Lewandowski et al.]
{Wojciech Lewandowski,$^1$\thanks{E-mail: boe@astro.ia.uz.zgora.pl}
Karolina Ro\.zko,$^1$
Jaros{\l}aw Kijak,$^1$
Bhaswati Bhattacharyya,$^2$\newauthor
Jayanta Roy,$^{2,3}$\\
$^1$Institute of Astronomy, University of Zielona G\'ora , Szafrana~2, 65-246~Zielona~G\'ora, Poland,\\
$^2$Jodrell Bank Centre for Astrophysics, School of Physics and Astronomy, University of Manchester, M13 9PL, UK,\\
$^3$National Centre for Radio Astrophysics, Tata Institute of Fundamental Research, Pune 411 007, India}
\date{Accepted . Received ; in original form }
\maketitle

\begin{abstract}

We present the results of the multi-frequency scatter time measurements for ten radio pulsars that were relatively less studied in this regard. The observations were performed using the Giant Meterwave Radio Telescope at the observing frequencies of 150, 235, 325, 610 and 1060~MHz. The data we collected, in conjunction with the results from other frequencies published earlier, allowed us to estimate the scatter time frequency scaling indices for eight of these sources. For  PSR~J1852$-$0635 it occurred that its profile undergoes a strong evolution with frequency, which makes the scatter time measurements difficult to perform, and for PSR~J1835$-$1020 we were able to obtain reliable pulse broadening estimates at only two frequencies.  
We used the eight frequency scaling indices to estimate both: the electron density fluctuation strengths along the respective lines-of-sight, and the  standardized amount of scattering at the frequency of 1~GHz. Combining the new data with the results published earlier by Lewandowski et al., we revisited the scaling index versus the dispersion measure (DM) relation, and similarly to some of the earlier studies - we show that the average value of the scaling index deviates from the theoretical predictions  for large DM pulsars, however it reaches the magnitude claimed by L{\"o}hmer et al. only for  pulsars with very large DMs ($>$650~pc~cm$^{-3}$). We also investigated the dependence of the scattering strength indicators on the pulsar distance, DM, and the position of the source in the Milky Way Galaxy.

\end{abstract}

\label{firstpage}


\section{Introduction}
 \label{intro}

The phenomenon of interstellar scattering of pulsar radio signals has been extensively studied for almost 50 years since it was identified \citep{scheuer68}. In our recent studies that were presented in \citet[hereafter Paper~1]{lewan13} and \citet[Paper~2]{lewan15}, we investigated the multi-frequency properties of the scattering phenomenon, following the earlier works of \citep[hereafter L01 and L04 respectively]{L01,L04}. The special aim of these studies was the estimation of the scatter time frequency scaling index $\alpha$ which is one of the crucial scattering parameters that - theoretically - allows us to deduce the turbulent properties of the interstellar medium (ISM).

In Paper~2 we analysed the multi-frequency scatter time measurements for 60 pulsars, but, at the same time, we noticed a large number of sources for which there were only one or two odd measurements (in the frequency range from 150~MHz to  1~GHz) indicating a significant scattering. While two measurements are too few to reliably estimate the frequency scaling of the pulsar's scattering and to include it in our multi-frequency analysis, we found such sources to be excellent candidates for the next observational project using the Giant Meterwave Radio Telescope (GMRT, located near Pune, India) - an interferometer that can be easily used in the phased array mode allowing for observations in five frequency bands: 150, 235, 325, 610 MHz and 1.0 to 1.4 GHz (the L-band receiver).

Multi-frequency observations of the properties of scattered pulsar profiles are crucial to our understanding of the phenomenon. The current scattering theory (see \citealt{rickett90} for a review; also see \citealt{rama97}, \citealt{lambert99}, \citealt{cordes01}, \citealt{bhat04}, L04, \citealt{rickett09}, \citealt{brisken10}) assumes that the pulsar profiles are broadened due to the different lengths and travel times of the pulsar rays that are scattered in the ISM. As a result, the pulsar profiles attain the so called ``scattering tails''. The shape of the tail depends on the geometry of scattering. In the simplest case, we can assume that all of scattering occurs in a thin screen located somewhere between the pulsar and the observer. If we assume that the distribution of the scattering angles takes a form of a two-dimensional gaussian function, then the observed pulsar profile can be explained as a convolution of the intrinsic pulsar profile with a simple exponential decay function (often called the pulse broadening function, PBF). The characteristic decay time of this function ($\tau_d$) is called the scatter time or pulse broadening time.

The actual brightness distribution of the scattering can be estimated using a model of the turbulence in the ISM. In a homogeneous and isotropic medium the spectrum of the spatial electron density can be approximated as \citep{rickett77}:

\begin{equation}
\label{elec_dens_simple}
P_{n_e} (q) = C_{n_e}^2 q^{-\beta},
\end{equation}

\noindent 
where it is assumed that the scale of the fluctuation $q$ causing the scattering is neither close to {\it inner scale} nor to {\it outer scale} of the turbulence spectrum, and $C_{n_e}$ is the density fluctuation strength. It was shown that the fluctuation spectral index $\beta$ has to be lower than 4.0 
\citep{romani86}, and  the scatter time will depend on the observing frequency according to a power-law: $\tau_d \propto \nu_{obs}^{-\alpha}$. The observed scatter time frequency scaling index $\alpha$ can be bound to the spectral index of the fluctuation spectrum by a relation: $\alpha = 2\beta/(\beta -2)$. For a Kolmogorov's spectrum of the density fluctuations  ($\beta = 11/3$), the expected scatter time scaling index is $\alpha=4.4$, and the lowest allowable value for the thin screen scattering geometry is $\alpha=4.0$, which occurs for the so called critical model (with $\beta=4.0$)

The pulse broadening spectral slope is the same as the frequency scaling index of the decorrelation bandwidth in the interstellar scintillation theory, since both the scatter time and the decorrelation bandwidth $ \Delta\nu_d$are bound by the relation:

\begin{equation}
\label{def_c}
2\pi\, \tau_d\, \Delta\nu_d = C_1.
\end{equation}

\noindent
In this formula $C_1$ is constant and presumably close to unity, however, this value will slightly differ for different turbulence models and scattering geometries \citep{lambert99}. The scatter time measurements and scintillation based measurements can be used to ascertain the value of the scaling index which is especially useful for nearby pulsars, where the scatter time is usually very small (unless the observations are conducted at very low frequencies).
 
The multi-frequency pulse broadening and/or decorrelation bandwidth analysis was attempted in the past for a limited number of pulsars. \citet{cordes85} analysed data for 76 sources, however, only for five of them they obtained measurements on three or more frequencies. Similar data were also published by \citet{johnston98} with another five sources. L01 and L04 collected that data, along with publications by other authors, and added their own pulse broadening measurements to obtain $\alpha$ estimates for 27 sources. In Paper~1 and Paper~2, we increased this number to 60, still this data set may be considered small when compared to some other scattering (and non-multi frequency) analyses: for example \cite{bhat04}, who observed 98 pulsars and gathered 371 measurements, or the recent work of \citet{krishna15}, who measured scatter times for 128 sources, then, for the statistical study, analysed 385 single frequency measurements. One has to remember, however, that multi-frequency measurements are more useful in scattering studies, since the interpretation of the results does not require invoking the scattering theory (nor the simplified geometry models), as often as it takes place in the case of single frequency studies. Therefore, we believe that increasing the number of objects with multi-frequency pulse broadening measurements will be crucial for our understanding of the scattering phenomenon itself, as well as the properties of the Galactic  ISM causing it - both physical and geometrical. To achieve that, we started an observing campaign using the GMRT at all available observing bands. In this paper we report the first stage of the project which involved observations of ten pulsars.

\section{Observations and Analysis}
\label{obs}

We selected ten pulsars which are relatively strong and have shown significant scattering in the 150~to 1000~MHz frequency range in the earlier observations, while at the same time they did not have a satisfactory frequency coverage for pulse broadening measurements - usually just one or two odd measurements, which did not allow for a reliable scatter time frequency scaling analysis. Another important selection criterion was that the pulsar has a relatively simple average profile (at high observing frequencies where there is no scattering present) with no apparent asymmetries. We wanted to avoid multiple-component profiles, and especially the profiles containing two or more components with similar strength and - if that was the case - we needed the components to be widely separated (i.e. the separation of the components had to be larger than the expected scatter pulse broadening). These criteria were aimed at both: the simplification of the data analysis and increased reliability of our results, as they helped us to alleviate some of the possible sources of the erroneous scatter time measurements (see the discussion in Paper~1 and Paper~2). Apart from the pre-selected sample of about 20 sources, we finally decided on observing ten pulsars that are  located relatively close in the sky which ensured the most effective use of the telescope time (see Table~\ref{tau_measurements} for the list of sources).

The observations were conducted between May 09 and May 23 2014 using the Giant Meterwave Radio Telescope (GMRT, Pune, India) in its phased array mode. We observed in all the frequency bands available, and, depending on the expected amount of scattering, each pulsar was observed at three or four frequencies. The highest frequency each source  was observed at, was chosen in such a way, that it should not exhibit any measurable scatter broadening, and we used the profile shape from this frequency as a reference profile in our analysis. 

Data collection was made using the GMRT Software Backend \citep[{\sc GSB},][]{roy2010}, using the coherent array mode of about 60\% of GMRT antennas (the Core and 2 to 3 antennas in the interferometer ``arms''). The observations at each frequency were performed as a separate observing session. At the frequencies of 150~MHz and 235 MHz, we collected the data using  16~MHz bandwidth with 256 spectral channels and 112~$\mu$s sampling. For the 325~MHz, 610~MHz and 1060~MHz, we used 32~MHz bandwidth with 512 spectral channels and 243~$\mu$s sampling.  The typical integration time was 10 minutes, and only a few selected (weaker) sources were observed for a maximum of 30 minutes. Data was then dedispersed and folded with the topocentric pulsar period, which was calculated using {\sc TEMPO}\footnote{\tt http://www.atnf.csiro.au/research/pulsar/tempo} software, and the pulsar ephemeris from the ATNF PulsarCatalogue\footnote{Available at {\tt http://www.atnf.csiro.au/people/pulsar/psrcat/}} \citep{manchester05}. Prior to the dedispersion routine, we also applied a robust procedure to clear the data of strong interference, by removing spectral channels heavily affected by narrow band interference and flagging periods of time affected by unwanted broadband signals.

 \begin{figure}
\resizebox{\hsize}{!}{\includegraphics{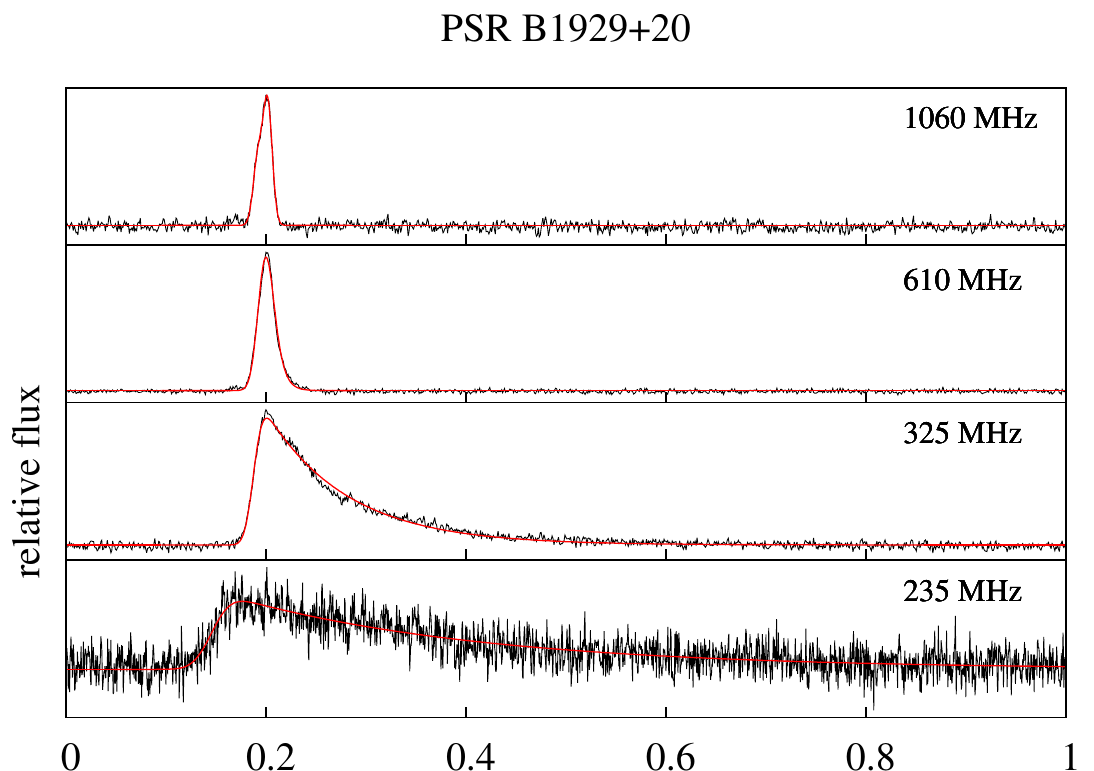}}
\vskip3mm
\resizebox{\hsize}{!}{\includegraphics{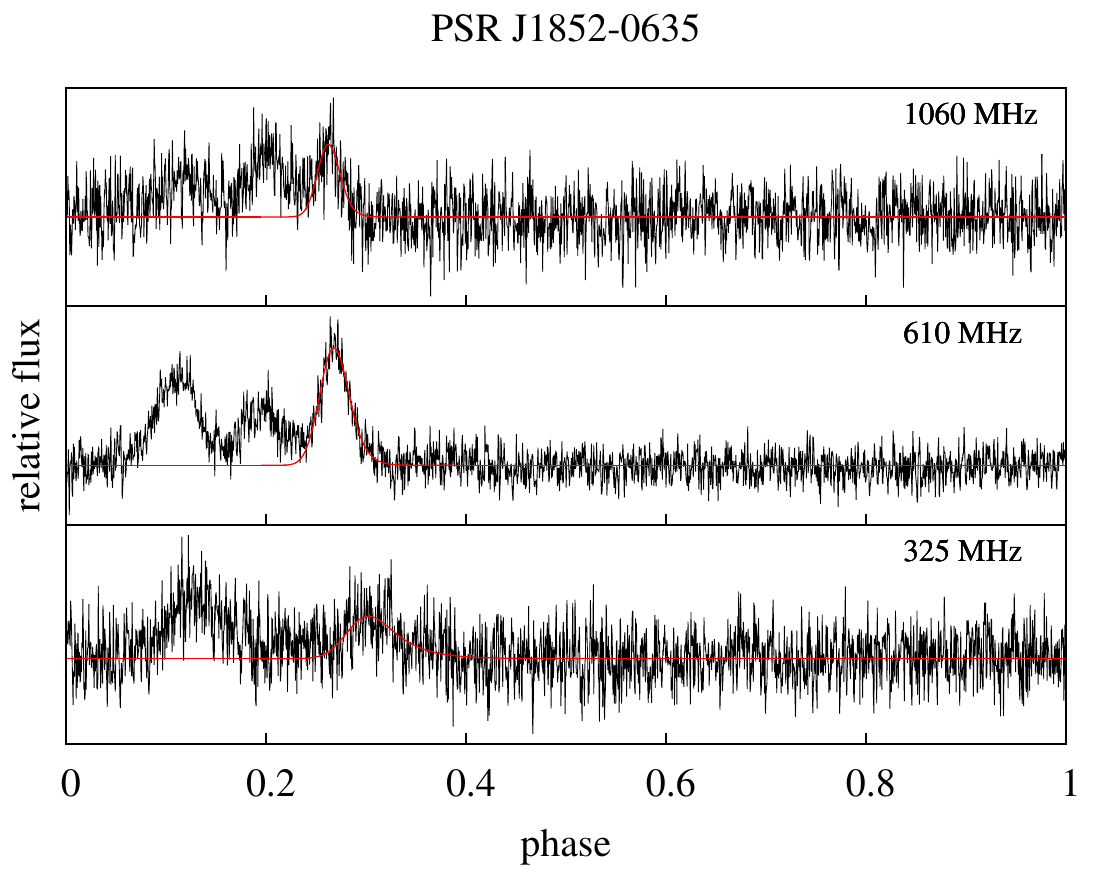}}
\caption{A sample of the profiles from our analysis. PSR~B1929+20 which was detected at four frequencies, PSR~J1852$-$0635 at three. Red lines overlapping the profiles show the result of  our fits. In case of the second pulsar, we based our analysis on modelling of only the rightmost component.
\label{sample_profile}}
\end{figure}

Two sample sets of pulsar profiles we obtained are shown in Figure~\ref{sample_profile}. The integration times we used were chosen to assure the signal-to-noise ratio of at least 50, however, due to a heavy interference present, especially at lower frequencies and possible underestimation of the sky background brightness (all the pulsars we observed are very close to the Galactic plane), we were not able to reach this goal. Few of the pulsars were not detected at all in the 150~MHz and 235~MHz data. This was partially due to the very strong scattering - like in cases of pulsars B1830$-$08 at 325~MHz or B1913+10 at 235~MHz.  In some other cases, the inherently low flux density of the pulsar was an issue, like for PSR~J1852$-$0635 which was in the meantime identified as a gigahertz-peaked spectrum source (see \citealt{dembska14}, and also \citealt{kijak11} for general information about GPS pulsars). Nevertheless, we managed to obtain 32 pulse profiles for 10 pulsars (see Figures~\ref{app_profiles} in the Appendix A) and successfully measure 26 pulse broadening times, which are presented in Table~\ref{tau_measurements} (the remaining six had very little scattering and were used mainly as profile templates; see the next section).

\subsection{Data Analysis and Results}

There are at least a few methods of finding the pulse broadening of pulsar signals. \citet{demorest11} proposed the cyclic spectral analysis data, which requires access to the raw voltages which were not available in our case. \citet{bhat03} proposed the ``CLEAN''  method which attempts to de-convolve the intrinsic pulsar profile from the scattering. We, however, opted for the same method we used in our previous analysis (see Paper~1 and Paper~2 and the detailed discussion there) which involves the fitting of a model (template) profile convolved with a pulse broadening function (PBF) to the observed profile by adjusting the amount of the pulse broadening. According to \cite{rama97}, the observed pulse profile can be represented by:

\begin{equation}
\label{profile_shape}
P^O(t) = P^I(t) *  s(t) *  d(t) * i(t),
\end{equation}

\noindent
where $P^I(t)$ is the intrinsic profile shape,  $s(t)$ is the scattering smearing function in the ISM, $d(t)$ is the ISM dispersion smearing function,  $i(t)$ is the receiving backend response function, and asterisks ($*$) denote a convolution. In case of our data, the dispersion smearing and the backend response functions have only a negligibly small influence on the observed profile. 

\begin{table*}
\caption{Measured scatter time values. The dash (``-'') denotes that the pulsar was not attempted/not 
detected in our data, and zero value indicates a detection of the pulsar but no measurable pulse broadening (using 3-$\sigma$ threshold). The uncertainties given in the table are 3-$\sigma$ values from our fits.\label{tau_measurements}}
\begin{tabular}{llllll}
\hline
Pulsar & \multicolumn{5}{c}{$\tau_d$ (ms)}\\
&  150 MHz & 235 MHZ & 325 MHz & 610MHz & 1060 MHz \\ \hline
B1620$-$42    &-   &-   & 75.7$\pm$3.4 &9.2$\pm$0.4 & 0.65$\pm$0.30 \\
B1830$-$08    & -  &-   & -   &31.2$\pm$2.3 & 2.8$\pm$0.3\\
J1835$-$1020 &-   &-   &17.9$\pm$7.3 & 2.57$\pm$0.54 & 1.56$\pm$0.83\\
J1852$-$0635 &-   &-   &14.2$\pm$4.9 & 4.8$\pm$0.9 &  3.1$\pm$3.0 \\
B1911$-$04    & 19.9$\pm$0.8 & 2.80$\pm$0.07& 0 & 0 & - \\
B1913+10       & -  &-   & 63.4$\pm$5.6 & 4.83$\pm$0.21 & 1.90$\pm$0.35 \\
B1914+13       &-    & 17.5$\pm$3.6 & 7.64$\pm$0.61 & 1.30$\pm$0.38& -\\
B1929+20       & - & 68.3$\pm$5.0 & 21.8$\pm$0.5 & 1.71$\pm$0.21& 0\\
B1953+50       & 2.77$\pm$0.80 & 0.506$\pm$0.003 & 0 & 0&- \\
B2002+31       & -  & 45.0$\pm$4.9& 11.14$\pm$0.35& 0&- \\
 \hline
\end{tabular}
\end{table*}

The intrinsic pulsar profile is always unknown, and there are a few ways to model its shape. L01 and L04 and, recently, \citet{krishna15} used a method in which a profile from a higher observing frequency (i.e. high enough that the scatter broadening of the profile can be neglected) was used as a template. This method can not, however, take into account the
possibility of the frequency evolution of the profile. Therefore, in our analysis we used the same method as in our Paper~2 (and to some degree in Paper~1). To model the intrinsic pulse shape we used a gaussian function or a sum of up to three gaussians. This template profile was obtained by fitting a linear combination of gaussian functions to the profile obtained at the highest frequency, the pulsar was observed (which was presumably not affected by scattering).
In the next steps, for the template profile, the relative amplitudes and distances of the gaussians were kept fixed, however, to take the various effects of the pulse shape evolution into account  - such as the radius-to-frequency mapping \citep[see][]{kijak03} - we allowed for the widths of the gaussian function(s) to vary within reasonable limits. In practice we had to allow the widths to increase (by a factor of 3 maximum) only for the lowest frequency and most scattered profiles, where the increase of the width was negligibly small compared to the measured value of the scatter time, and we can safely assume that any errors made by using that approach are included in the total uncertainty of the fit (which following L01, \citealt{krishna15} and others we assumed to be three times greater than the statistical fit error). The scattered profiles were then modelled by a convolution of the template with the pulse broadening function $PBF_1$, which represents scattering on the thin screen \citep[see][]{williamson72, williamson73} , and may be expressed as (see for example L04):

\begin{equation}
\label{pbf1}
PBF_1(t) = \exp(-t/\tau_d) U(t),
\end{equation}

\noindent
where $U(t)$ is the unit step function ($U(t<0)=0$, and $U(t \geq 0) = 1$).
 
The scatter time values were found using the least squares fit of the modelled profile to the observational data, and
were performed using the {\sc Origin} software\footnote{http://www.originlab.com/}. Table~\ref{tau_measurements} shows the results of our measurements. The dashes indicate that the pulsar was either not detected in our data, nor attempted at all at a given frequency (usually because the expected scattering was too large or too small to be measured). The zero values correspond to the observations in which the pulsar was observed, but we could not reliably measure the amount of scattering, i.e. the pulse broadening time our analysis yielded was smaller than the corresponding measurement uncertainty. The results of our fits are shown as red lines in Figure~\ref{sample_profile} and Figure~\ref{app_profiles} (in the Appendix A).

As we mentioned above (at the beginning of Section~\ref{obs}), our sample of pulsars was carefully selected to simplify the analysis and increase the reliability of our estimates. There should be no additional  sources of uncertainty of our measurements (such as the intrinsic profile asymmetry), but we can never exclude the possibility of significant intrinsic profile evolution of these pulsars. One special case in this regard is PSR~J1852-0635 (see the profiles in Figure~\ref{sample_profile}), whose profile significantly differs from frequency to frequency. It is impossible to tell if the reason for these variations is an evolution of the profile, or if we are dealing with a mode-changing pulsar - the latter cannot be excluded, since our profiles come from 10 to 30 minute observations and the pulsar was observed at each frequency at a different epoch. Nevertheless, we believe that even in this case our measurement may be considered reliable - the components of the profile are widely separated and in our analysis we modelled the scattering only for the right-most component.

\section{Results and discussion}
\label{section_evolution}

The scatter time frequency scaling index is the most obvious and useful outcome of multi-frequency scattering or scintillation studies. As we mentioned in Section~1 it gives us means to assess the spectrum of the turbulence of electron gas in the ISM. In a homogeneous and isotropic medium, one expects the frequency scaling index $\alpha$ to be between 4.0 and 4.4, depending on the ISM turbulence model \citep{romani86}. Any deviations from this may provide us with clues about the geometry of scattering  \citep[see ][]{cordes01}, or other properties of the turbulence spectrum (such as inner/outer scales, see the Introduction).

Estimation of pulse broadening at multiple frequencies for pulsars from our sample allows us to perform such studies for these objects. These pulsars were specifically picked - amongst other criteria - for having had a relatively poor, to virtually non-existent, frequency coverage of scatter time measurements until now. Because of this, the results of our analysis provide first reliable estimates of the frequency scaling indices for these pulsars, or at least represent a vast improvement over the earlier estimates.

\begin{figure*}
\resizebox{\hsize}{!}{\includegraphics{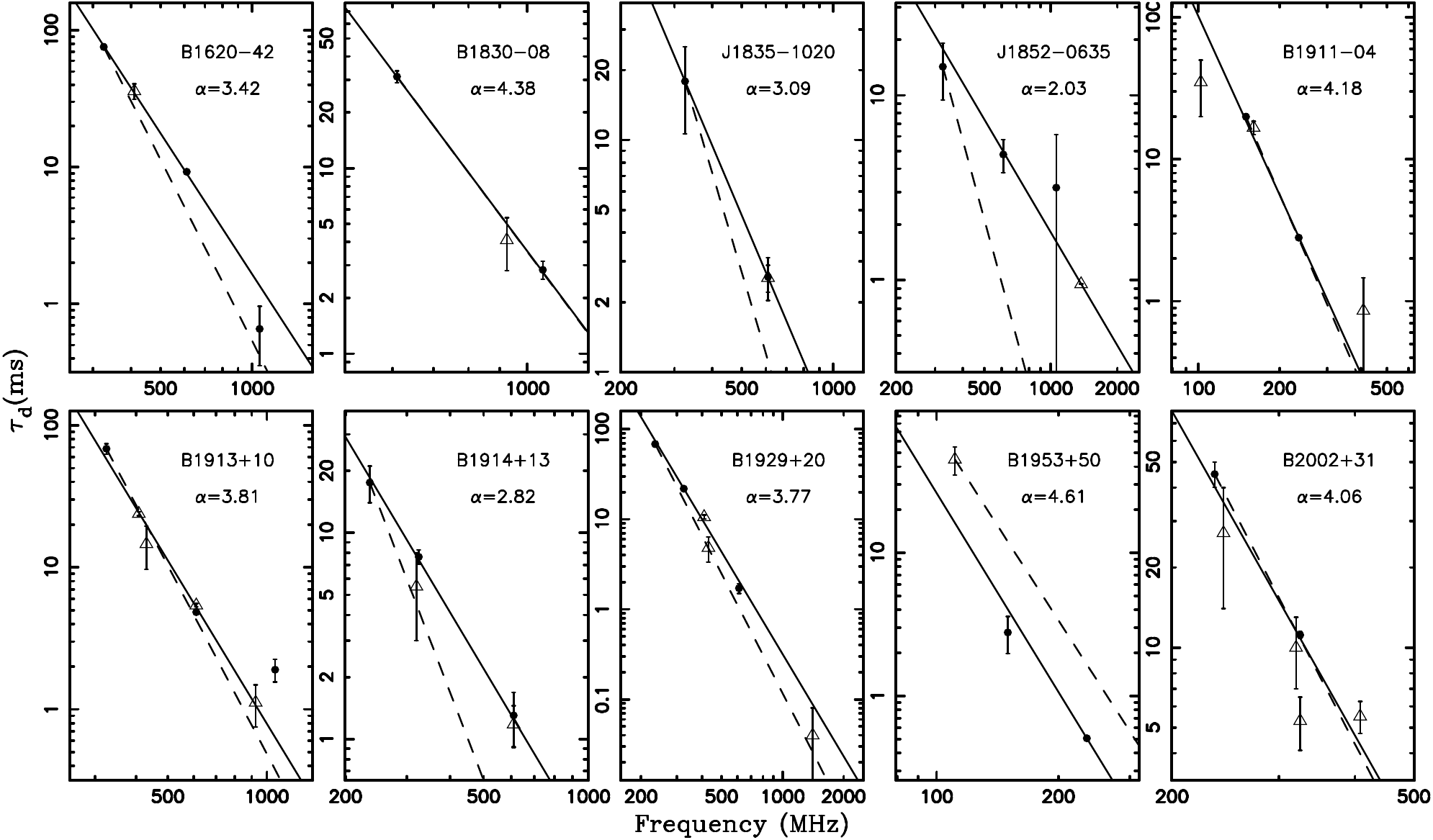}}
\caption{Scatter time versus the observing frequency for ten pulsars from our sample. 
The solid line in each plot is the representation of the power-law fit to the scatter time measurements, while the dashed line shows the slope of the thin-screen Kolmogorov prediction with $\alpha = 4.4$. 
\label{plot_tau_vs_freq}}
\end{figure*}

To make our study more complete,  we supplemented our data with the archival scatter time estimates for these pulsars - see Paper~1 and Paper~2 (and references therein). 
From all the collected data we removed estimates that were clearly erroneous, which were usually the estimates at frequencies above 1~GHz or some of the oldest estimates at very low frequencies - see the discussion in Paper~2. The remaining data are shown in Figure~\ref{plot_tau_vs_freq}, where our new estimates are represented by dots, and the measurements published earlier are represented by triangles\footnote{for full reference list see the Table in the Appendix of Paper~2, and for the  B1830$-$08 and J1852$-$0635 data see Table~1 in Paper~1.} . As one can see our scatter time measurements are in a relatively good agreement with the earlier estimates.
 
We estimated the scatter time scaling indices by modelling the $\tau_d$ versus frequency data with a power-law function:

 \begin{equation}
 \label{fit_form}
 \tau_d(\nu)= 10^{-\alpha \log \nu +b},
\end{equation} 
 \noindent
(where $b$ is a free parameter that will be used later for the estimates of the scattering strength) by means of weighted least squares method, where the weight of each scatter time estimate was the inverse square of its uncertainty. The results of our fits are shown as solid lines in Figure~\ref{plot_tau_vs_freq}. In some cases (like for PSR~B1911$-$04), it may appear as if some of the data points were omitted, this is however not the case. The apparent omission is only a result of relatively large uncertainties of these data which greatly reduced their weights in our fits. 

To estimate the uncertainty of the resulting $\alpha$ we initially employed the method of $\chi^2$ mapping, however in some cases it yielded unbelievably small error estimates. This was for pulsars with only three measurements and a very small data spread, in which case the fit modelled the data almost perfectly, yielding very small value of minimum $\chi^2$ despite the large uncertainties of individual scatter time measurements. An example of such  dataset are the measurements for PSR J1835$-$1020 (see Fig.~\ref{plot_tau_vs_freq}). To overcome this we employed an alternative method using Monte-Carlo simulations to evaluate uncertainties. We used the larger of the two estimates  as the final uncertainty of $\alpha$.

\begin{table*}
\caption{The scattering spectral index ($\alpha$) and the electron density spectral index ($\beta$) for the observed pulsars. Values of $\alpha$ quoted in italic is  considered doubtful (see article text for explanation). The $\alpha$ fit uncertainties were calculated using $\chi^2$ mapping method or the Monte-Carlo method (the latter are marked by $^{MC}$, see article text for explanation). Table also lists the scattering fluctuation strength  $\log C_{n_e}^2 $ and the estimated scattering at a standard frequency of 1~GHz. \label{table_alpha}}
\begin{tabular}{lrrrrllcc}\hline
Pulsar & $l_{II}~~$ &  $b_{II}~~$ & $DM$~~~~~ & Distance & $\alpha$~~ & $\beta$~~ & $\tau_d$ (ms) & $\log[ C_{n_e}^2(\mbox{m}^{-20/3})] $\\
      & (deg) & (deg) & (pc cm$^{-3}$) & (kpc)~~~ & & &(at 1 GHz) &\\ \hline \hline
B1620-42&    338.89  &    4.62   &     295(5)      &  6.58  &   3.42$\pm$0.21& *  & 1.65&$-$1.25\vspace*{0.5mm}\\
B1830$-$08&  23.39   & 0.06    &  411(2)     & 4.65    &   4.38$\pm0.35^{MC} $   &   3.7$\pm$0.6   & 3.65  &$-$1.61\vspace*{0.5mm}\\   
 J1835$-$1020&  21.98   & -1.30   & 113.7(9)    & 2.30   &   {\it 3.09$^{+0.83}_{-0.96}\phantom{i}^{MC}$}  &  *     & *& *\vspace*{0.5mm}\\ 
 J1852-0635&    27.22   &  -3.34    &    171(6)      & 4.00   &  {\it2.03$\pm$0.95}$ ^{MC}$   &    *   & * &*\vspace*{0.5mm}\\
B1911$-$04&    31.31   &  -7.12    &     89.385(10)   & 2.79   &    4.18$^{+0.44}_{-0.41}$  &   3.83$\pm0.8$  & 0.0067&$-$3.64\vspace*{0.5mm}\\
B1913+10&    44.71   &  -0.65    &    241.693(10)   &  7.00   &   3.81$^{+0.43}_{-0.39}$ &    4.2$^{+1.0}_{-0.9}$    & 0.79&$-$2.15\vspace*{0.5mm}\\
B1914+13&     47.58  &    0.45   &     237.009(11)  &   4.50  &    2.82$\pm$0.26$ ^{MC}$  &   *    &  0.31&$-$2.49\vspace*{0.5mm} \\
B1929+20&    55.57   &   0.64    &    211.151(11)   & 5.00   &   3.76$^{+0.36}_{-0.42}$   &    3.6$^{+0.8}_{-0.9}$       & 0.32& $-$2.26\vspace*{0.5mm}\\
B1953+50&     84.79  &   11.55   &      31.974(3)  &  2.24 &     4.61$^{+0.58}_{-0.66} \phantom{i}^{MC}$  &    3.5$^{+0.9}_{-1.0}$  & 0.00063 & $-$3.91\vspace*{0.5mm}\\
B2002+31&    69.01   &   0.02    &    234.820(8)   & 8.00   &   4.06$\pm$0.85   &    3.9$\pm$1.6      &0.11 &$-$3.26\vspace*{0.5mm}\\
    \hline
\end{tabular}
\end{table*}

Table~\ref{table_alpha} summarizes the results of our scatter time frequency dependence models, along with the relevant parameters of these pulsars, such as the galactic coordinates, distances and DMs (taken from ATNF Pulsar Catalogue, \citealt{manchester05}). For pulsars with $\alpha\geq4.0$ (within error estimates), we also calculated the resulting spatial electron density spectrum index $\beta=2\alpha/(\alpha-2)$, this relation does not apply, however, in cases where $\alpha$ is lower than 4.0. The table also shows two additional parameters: the normalized scatter time (at the observing frequency of 1~GHz) and the scattering strength parameter $\log C^2_{n_e}$. The normalized scatter time was calculated by extrapolating the power-law frequency dependence that we modelled for a given source to a frequency of 1~GHz. Using this value, we estimated the corresponding decorrelation bandwidth at this observing frequency (see Equation~\ref{def_c}), and the scattering strength was estimated by based on \citet{cordes85}:

\begin{equation}
\label{eq_c2}
C^2_{n_e}(\mbox{m}^{-20/3}) = 0.002 \ \nu^{\ \beta} \ D^{-\beta/2} \ \nu_d^{-(\beta-2)/2},
\end{equation} 

\noindent
where $\nu$ is the observing frequency in GHz (in our case $\nu=1$), $D$ is the pulsar distance (in kpc) and $\nu_d$ is the decorrelation bandwidth (in MHz). For the purposes of our calculations, we used the actual values of $\beta$ in cases where we
were able to calculate them, and the Kolmogorov's value of $\beta=11/3$ otherwise.

The values of the scattering scaling index we obtained range from $\alpha=4.61$ for PSR~B1953+50 down to $\alpha=2.03$ for J1852$-$0635. This last value, however, we consider to be unreliable. The amount of scattering visible in the profiles of this source (see Figure~\ref{sample_profile}), even at the lowest frequency we were able to estimate (i,e, at 325~MHz), is relatively small and due to the problems with the profile mentioned above, and rather poor signal-to-noise ratio, this measurement comes with about 30\% fractional error. We find it likely that all the measurements at higher frequencies, including the 1.3 GHz measurement reported in Paper~2 (which we show as a triangle in the respective panel in Figure~\ref{plot_tau_vs_freq}) can be considered only to represent upper limits for the actual scattering, as we suspect that 
our fits for the $\tau_d$ value probably misinterpreted the residual asymmetry of the profile as the scattering. For this pulsar, we can conclude, that finding the real scatter time frequency slope will be extremely hard, as it would require reliable $\tau_d$ estimates at frequencies below 325~MHz. This in turn would require a considerable amount of integration time on this gigahertz-peaked spectrum pulsar \citep{dembska14}; we attempted to detect this source at both 150~MHz and 235~MHz in 30 minute integrations with no success.  For similar reasons we rejected the 1060~MHz measurement for  PSR ~J1835$-$1020, however in this case it means that we have measurements at only two frequencies, which does not fulfil our requirement for multi-frequency measurement, and hence we excluded this pulsar from further analysis as well.

\begin{figure}
\resizebox{\hsize}{!}{\includegraphics{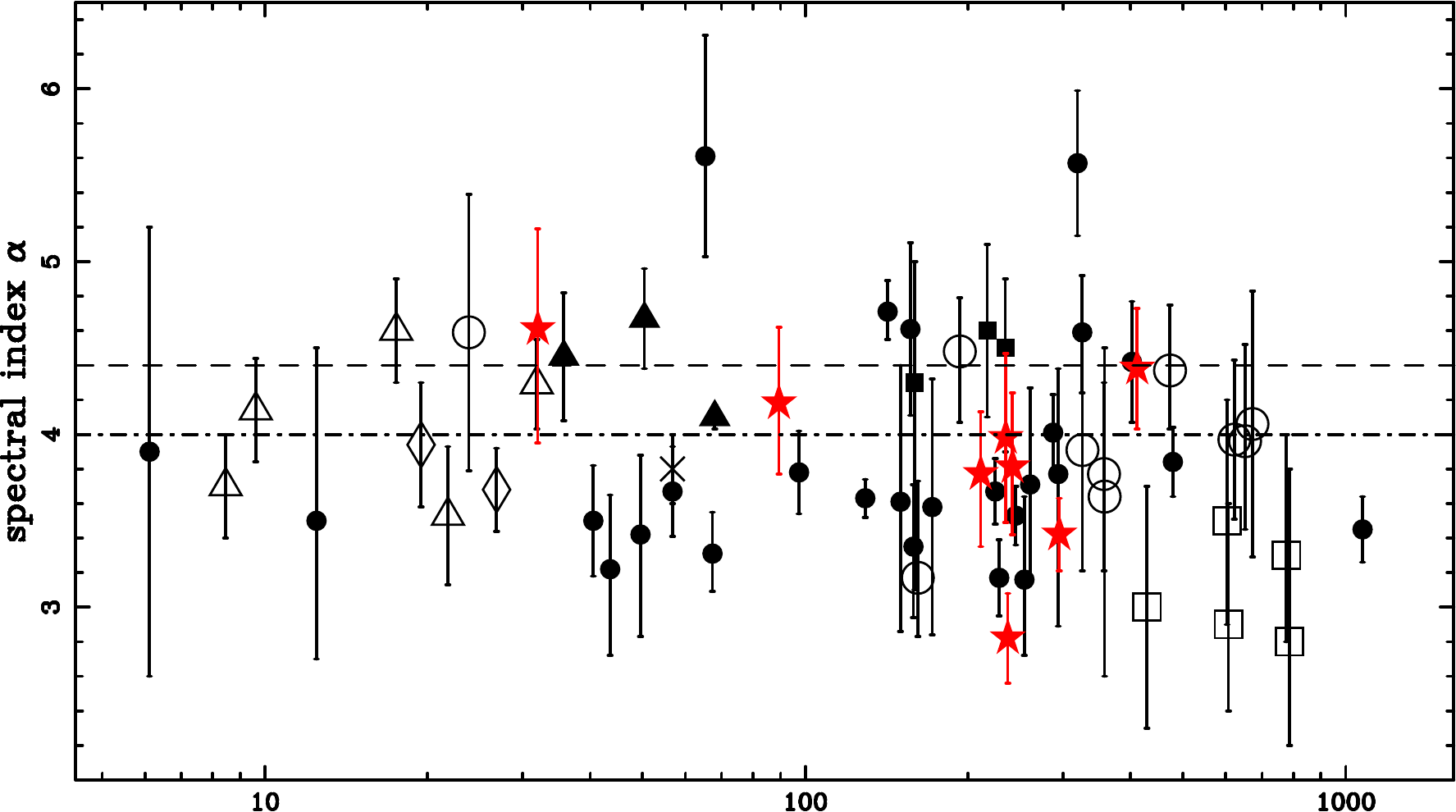}}
\vskip1mm
\resizebox{\hsize}{!}{\includegraphics{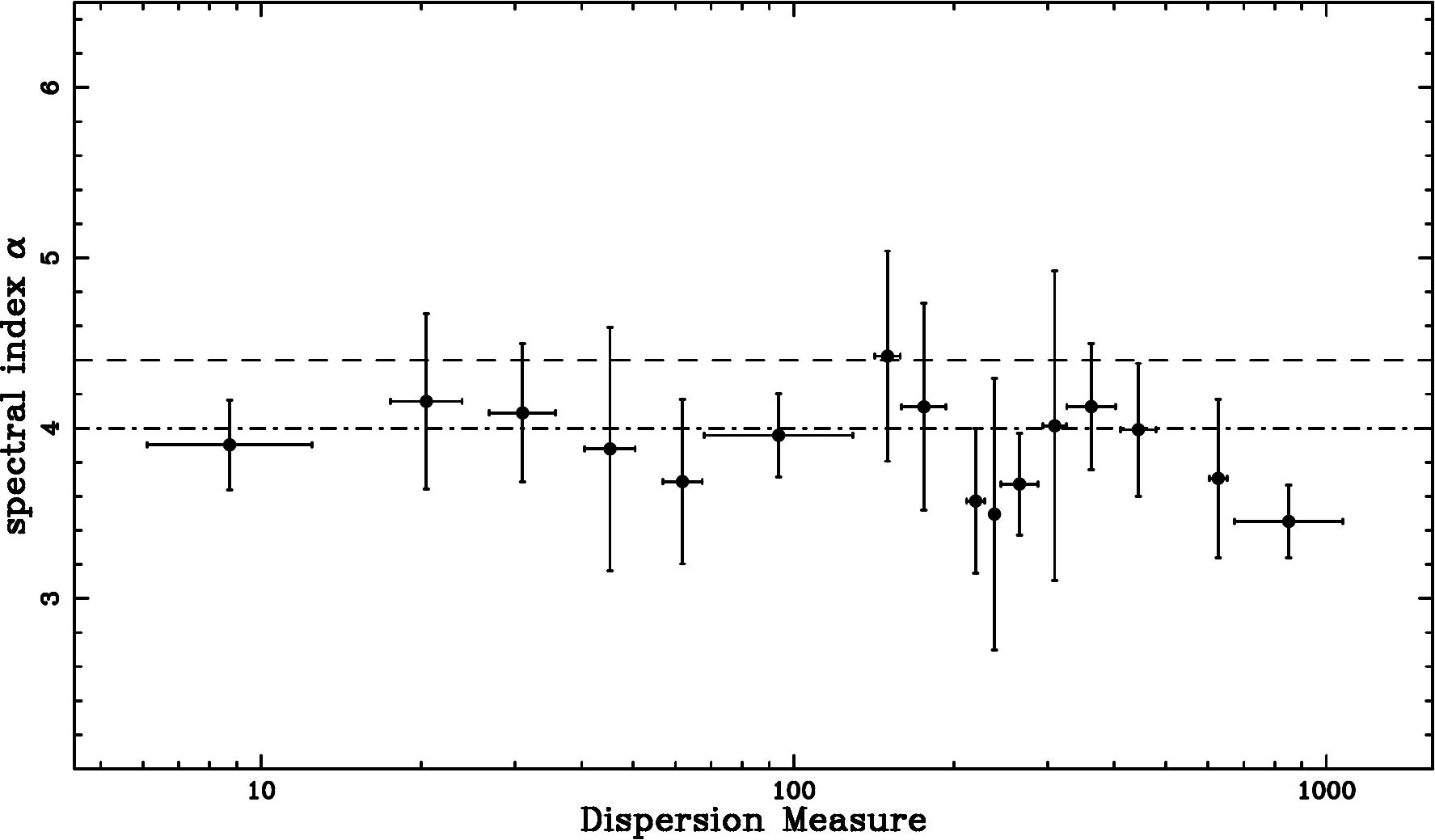}}
\caption{A plot of the spectral index of pulse broadening $\alpha$ versus the dispersion measure. On top on the top panel the empty squares represent the measurements from L01, filled squares - L04, empty triangles - \citet{johnston98}, filled triangles - \citet{cordes85}, cross - \citet{kuzmin02}, diamond - \citet{lewan11, daszuta13}, open circles  - \citet{lewan13}, full circles - \citet{lewan15}, stars (red) - current analysis.  The bottom panel shows the weighted average values of $\alpha$ binned in such a way that each DM-bin contains the estimates for four pulsars.\label{alpha_dm}}
\end{figure}

Disregarding the values of $\alpha$ for the two pulsars mentioned above, the value of the lowest pulse broadening frequency slope we found is $\alpha=2.82\pm0.26$ for PSR~B1914+13. This value is still far from the range allowed by the simple scattering geometries ($4.0\leq\alpha\leq 4.4$), however much closer to that range than it was estimated before ($\alpha=1.43\pm0.22$ in Paper~2, where it was deemed to be unreliable and was not used in later analysis). Still, we believe that the new estimate is solid, and the reason for the deviation may be either in the geometry of scattering or the turbulence spectrum in the scattering screen for this particular case (see the more detailed discussion in Paper~1). It is also worth mentioning, that similar values are observed for other pulsars as well. One of such objects is the well studied case of PSR~B1933+16 ($\alpha=3.35\pm0.15$ in Paper~2, see also L04), and coincidently, on the sky, this pulsar is located just a few degrees away from PSR~B1914+13.

Similarly for PSR~B1830$-$08 in Paper~1 we quoted $\alpha=2.13\pm0.32$ while our current estimate is $\alpha=4.38\pm0.35$, i.e. almost equal to the canonical prediction of the thin screen model with Kolmogorov's turbulence. This is no surprise, as the earlier estimate was based on just three doubtful high frequency scatter time estimates, and was regarded as unreliable in Paper~1. 
The single change in the opposite direction, i.e. away from the theoretically predicted range, is the case of PSR~B1929+20, for which in Paper~2 we reported $\alpha=4.53\pm0.65$ while after adding the new estimates the value of $\alpha$ is $3.76^{+0.36}_{-0.42}$. We have to point out however that the earlier estimate was based on just three data points, two of which were measurements at very similar frequencies (410 and 430 MHz), while in our current project we added measurements at 235, 325 and 610~MHz, where the scattering is large enough to be measurable (see profiles in the Appendix A), which makes the new estimate much more reliable.

\begin{figure}
\resizebox{\hsize}{!}{\includegraphics{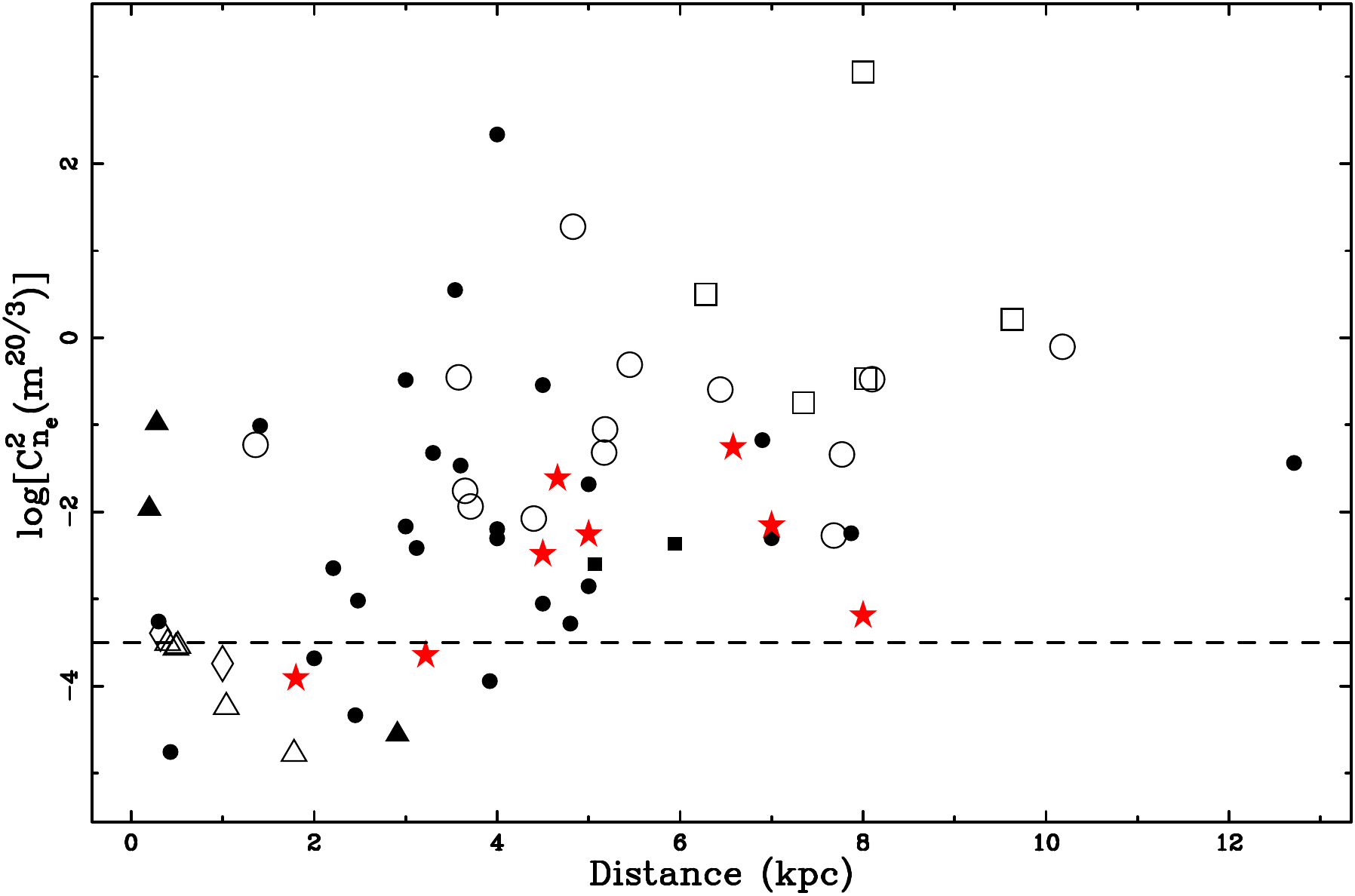}}
\caption{The fluctuation strength parameter $\log C^2_{n_e}$ plotted versus the pulsar distance. Symbols and colours used are the same as in Figure~\ref{alpha_dm}. \label{lc2_dist}}
\end{figure}

With the eight new or improved scatter time frequency slope estimates, and corrected $\alpha$ uncertainty estimates\footnote{a full set of corrected values of $\alpha$ and their uncertainties can be found in Table~\ref{tab3} in Appendix B. for all our fits (we employed the $\chi^2$ mapping plus Monte-Carlo simulations method described above to all the $\alpha$ estimates from the previous papers)}, we decided to revisit the $\alpha$ versus DM relation we investigated in Paper~1 and Paper~2 (following L01 and L04), and the result is shown in the top panel of Figure~\ref{alpha_dm}. The figure now includes the data for 64 sources and the new $\alpha$ measurements are denoted as red stars. 

The lower panel of Figure~\ref{alpha_dm} shows the weighted average values of the scatter time frequency scaling index, binned in such a way that each DM bin contains four pulsars. The horizontal bar represents the DM range over which the value was averaged, while the vertical bar represents the variance of the weighted average. As one can see the average values vary significantly from bin to bin. For DM's lower than 200~pc~cm$^{-3}$ they tend to stay relatively close to the range allowed by the simple scattering geometries, maybe except for the bin around DM=60~pc~cm$^{-3}$ which - amongst others - contains the Crab pulsar. For larger DMs the average $\alpha$ values seem to prefer much lower values (5 out of 8 bins in this range), however both the large variances of these averages (and the fact that three bins with large DM show  much higher average $\alpha$)) confirm a considerable spread of individual values clearly visible in the top panel of the Fig.~\ref{alpha_dm}. Also, one has to remember that while performing this kind of analysis we are averaging the scatter time frequency slopes of different objects, hence we are averaging the properties of scattering along vastly different lines of sight. These results should not be interpreted literally, for example as a way to predict the value of $\alpha$ for any given DM.

Nevertheless we find these data in quite a good agreement with findings from L01 and L04, with a possible slight correction, that even amongst the large-DM pulsars there is a considerable number of objects with values of the scatter time frequency scaling index in the 4.0 to 4.4 range, and that significant deviations from the simple geometry scattering predictions appear even for some of low-DM pulsars.

\begin{figure}
\resizebox{\hsize}{!}{\includegraphics{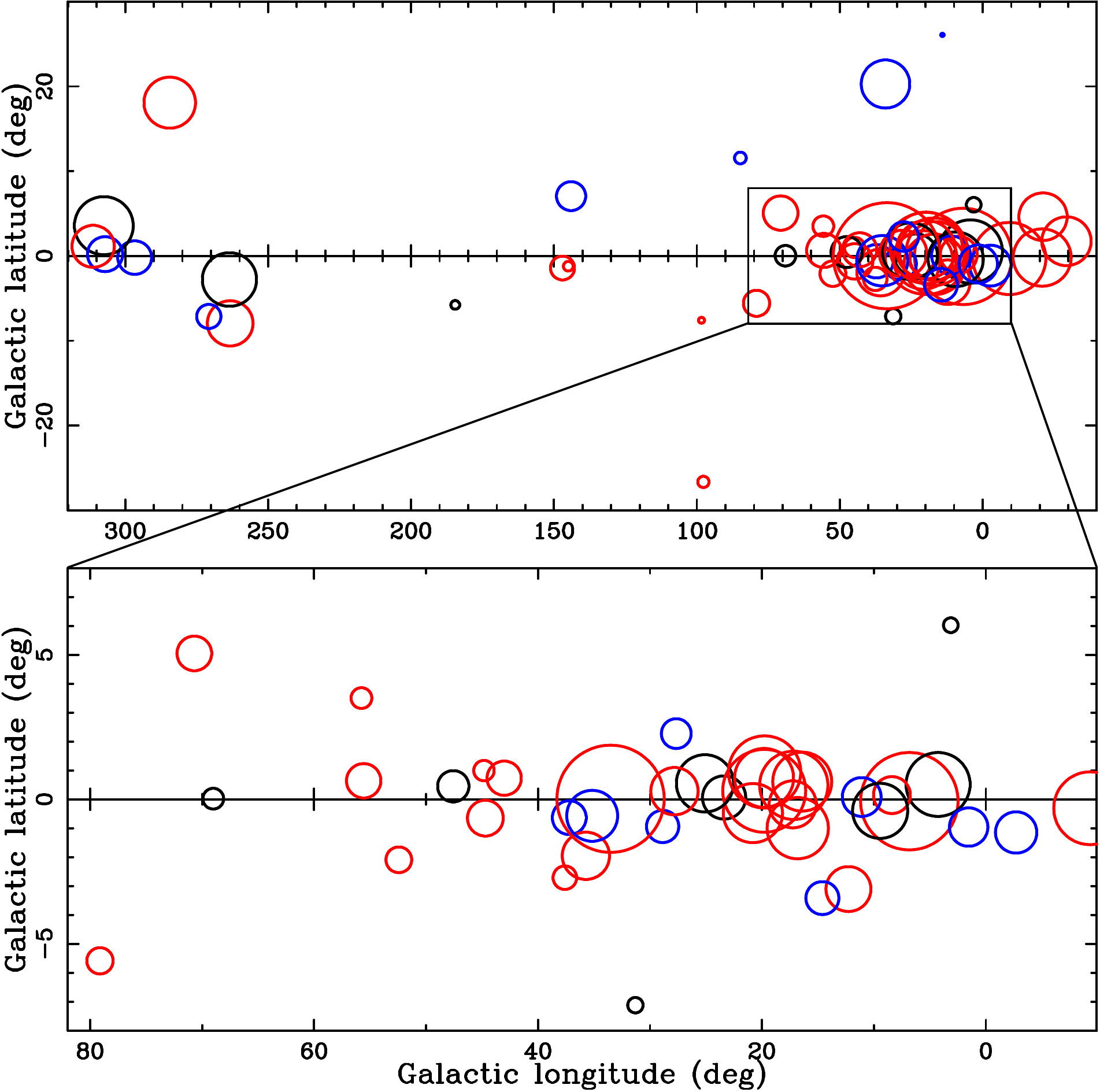}}
\caption{A plot of the fluctuation strength parameter $\log C^2_{n_e}$ in the galactic sky coordinates. The sizes of the circles correspond to the value of $\log C^2_{n_e}$, while the circle's colour indicates the value of the scatter time frequency scaling index $\alpha$: blue for $\alpha>4.4$, red for $\alpha<4.0$ and black for the remaining pulsars. The bottom plot shows a close-up of the Galactic Centre region\label{lc2_lb}}
\end{figure}

Figure~\ref{lc2_dist} shows a plot of the $\log C^2_{n_e}$ values against the pulsar distance for all the sources with multi-frequency scatter time measurements (i.e. from Paper~2 and the current work). As it was the case with the earlier plots of this quantity, which were based on single frequency measurements (the most recent was shown by \citealt{krishna15}), most of the values lie above the canonical prediction for the homogeneous Kolmogorov medium ($\log C^2_{n_e}=-3.5$, see \citealt{johnston98}). Our data show similar spread and trends as the previous analyses, one has to remember, however, that in the earlier attempts to calculate this value, the authors usually assumed pure Kolmogorov's spectrum, putting 
$\beta=11/3$  when using Equation~\ref{eq_c2}, while in our analysis we used the real values of $\alpha$ and $\beta$ (where applicable). Nevertheless, this did not influence the outcome in any dramatic way, as the change due to using our approach is usually negligibly small when compared with the spread of the data points for any given distance (which is clearly visible in Figure~\ref{lc2_dist}).

\begin{figure}
\resizebox{\hsize}{!}{\includegraphics{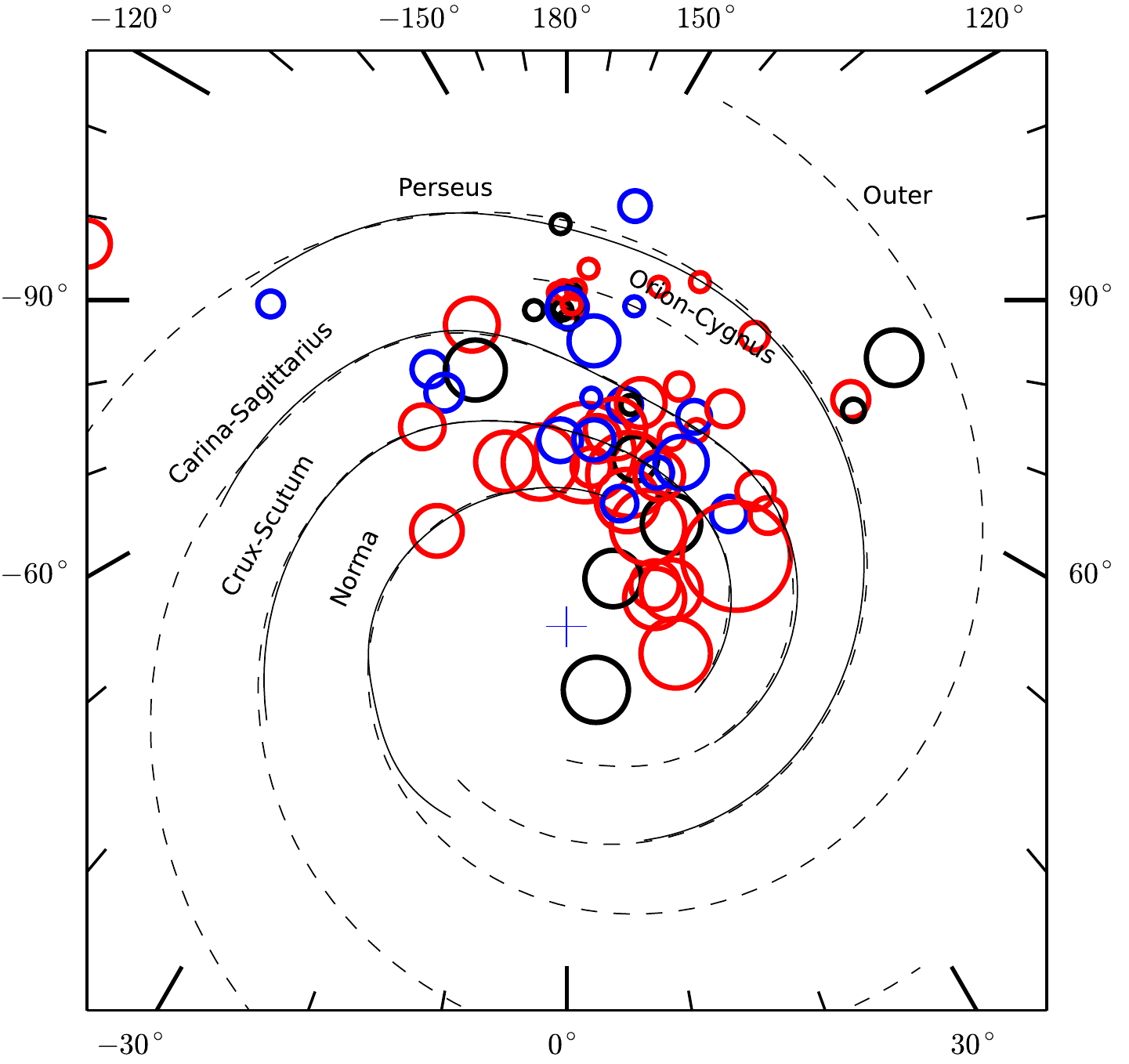}}
\caption{The fluctuation strength parameter for 64 pulsars in the Galactic X-Y coordinates, with the spiral arm positions from the NE2001 electron density model \citep{cordes2002}. The size of the circle indicates the value of $\log C^2_{n_e}$, while the colour denotes the value of the scatter time frequency scalling: red for $\alpha<4.0$, black for $4.0\leq\alpha\leq4.4$ and blue for $\alpha>4.4$. \label{lc2_xy}}
\end{figure}

We also plotted the value of $\log C^2_{n_e}$  in the Galactic sky coordinates (Figure~\ref{lc2_lb}) and in the Galaxtic X-Y coordinates (Figure~\ref{lc2_xy}). In these plots the size of the circle represents the magnitude of the fluctuation strength parameter, and the color indicates the value of the scatter time scaling index $\alpha$: blue for $\alpha>4.4$, black for $4.0\leq\alpha\leq4.4$, and red for $\alpha<4.0$. It is not surprising that the red circles dominate these plots (since the average value of $\alpha$ is just under 4.0), and, similarly to what we have shown in Paper~1 and Paper~2, there does not seem to be any correlation between the value of the scatter time scaling index and the position of the pulsar in the Galaxy. When it comes to the fluctuation strength $\log C^2_{n_e}$ obviously the larger values are preferred for distant pulsars, and the extremely large values can be seen only for sources that are apparently  behind the Carina-Sagittarius spiral arm, and in (or behind) the Crux-Scutum arm. One has to remember, however that the Galactic X-Y coordinates plot should not be over-interpreted, as  one of the crucial parameters used to create it - the pulsar distance - may be very inaccurate, as it is usually inferred from the dispersion measure.

One puzzling feature of Figure~\ref{lc2_lb} (bottom plot) may be the clustering of large red circles on the galactic plane near the galactic longitude of $l_{II}=20^{\circ}$. The number of measurements in this region may be easily explained: this region is still close to the Galactic Centre, which makes it densely populated by known pulsars. In the equatorial coordinates these pulsars lie in R.A. range between 18$^h$ and $18^h30^m$ and close enough to the celestial equator to make them easier to observe for the northern hemisphere radio telescopes, which explains the large number of multi-frequency pulse broadening measurements. This can not explain why all of the corresponding circles are red - i.e. why there are no sources in this direction showing $\alpha>4.0$. One has to remember, however, that all of the $\alpha$ estimates used to create this plot have some uncertainties and at least for some of them there is a chance that the actual scaling index is greater than 4.0, which would make this Galactic line-of-sight no different from the others. Hence, we believe that this feature may be just a simple coincidence.

\begin{figure}
\resizebox{\hsize}{!}{\includegraphics{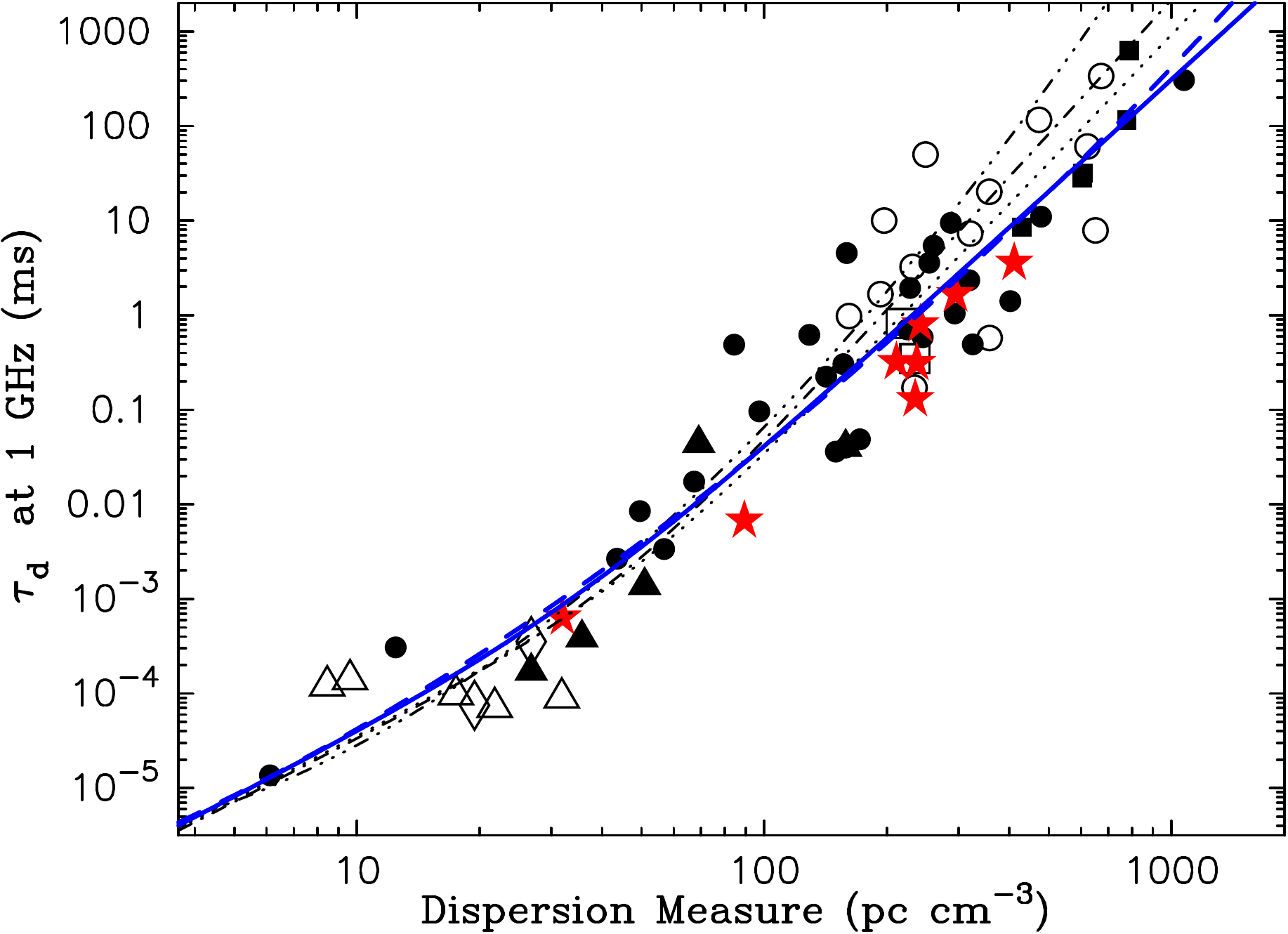}}
\caption{A plot of the 1~GHz scattering versus the dispersion measure for $C_1=5$.
The scattering-based estimates are shown by empty squares - L01, filled squares - L04, open circles - \citet{lewan13}, full circles - \citet{lewan15}, stars (red) - this paper.The  scintillation-based estimates are shown as: empty triangles - \citet{johnston98}, filled triangles - \citet{cordes85}, cross - \citet{kuzmin02}, and diamonds - \citet{lewan11, daszuta13}. The blue lines correspond to  $\tau$ versus DM relations proposed in Paper~2, and the previously proposed  dependences are shown by: the dash-dotted line for \citet{rama97}, the dotted line for L04 and the three-dot-dashed line for \citet{bhat04}. \label{bhat}}
\end{figure}

Figure~\ref{bhat} shows the plot of the amount of standardized scattering ($\tau_d$ at the frequency of 1~GHz) versus the dispersion measure for all the pulsars with multi-frequency scatter time measurements. Following Paper~2, for pulsars with the scintillation based estimates of $\tau_d$, we used $C_1=5$ to recalculate the decorrelation bandwidth measurements to pulse broadening times. The new and/or updated values obtained in the current analysis - represented by the red stars - seem to very well follow the relation(s) we adopted in Paper~2 (the solid blue lines).

\section{Conclusions}

In this paper we presented multi-frequency pulse broadening measurements for ten radio pulsars. For eight of them we were able - in connection with the previously published data - to obtain new estimates of the scatter time frequency scaling, which turned out to be a vast improvement over the earlier estimates for these sources. Using the new observational data in conjunction  with the results published in Paper~2, we investigated the dependence of the scattering parameters with the pulsar distance (and dispersion measure) and any possible correlations of these parameters with the pulsar's position in the Galaxy -  for a total of 64 objects with multi-frequency scatter time measurements\footnote{In Paper~2 we presented 60 estimates, while our current work updated four of them and added another four.}. This may be considered a small number compared to some other analyses - like the recent study of \citet{krishna15} which involved single frequency measurements for 385 pulsars. Yet, having multi-frequency information and the resulting frequency scaling index in our analysis, definitely helps to alleviate problems and biases with the scattering parameters - especially when it comes to re-scaling of these to some standard frequency (like the $\tau_d$ at 1~GHz), or removing the frequency dependence altogether (like for $\log C^2_{n_e}$). In case of our data set, we are able to use the frequency scaling that comes from the observational data of a given source, instead of using the theoretical predictions and/or values averaged for the whole population - as it was done in the past by
\citet{bhat04}, \citet{krishna15} and others.

  The average weigted value of the scatter time frequency scaling index is $\alpha=3.93\pm0.45$, where the uncertainty is $1\sigma$ standard deviation of the entire sample, hence it describes the spread of the individual measurements rather than the accuracy of the average. In the earlier studies of $\alpha$ versus DM dependence L01 and L04 suggested that for pulsars with DM$>$300~pc~cm$^{-3}$ the value of the scaling index $\alpha$ is significantly lower than for the low-DM pulsars.
The extension of the scaling indices data base we presented in Paper~2 (which we improved further in this work) allows us to modify this conclusion slightly. Firstly, deviations from the theory predicted range appear for selected pulsars of all DM ranges. Secondly, L04 claimed that the average scatter time frequency slope for pulsars with DM$>$300~pc~cm$^{-3}$ is much lower than predicted by simple geometry scattering theories, i.e. they obtained $\alpha=-3.44$ in this DM range. Using our database we are able to obtain such a low (weighted) average value only for the pulsars with DM$>$650~pc~cm$^{-3}$ ($\alpha=3.49\pm0.25$). At the same time the weighted average value for objects with DM$>$200~pc~cm$^{-3}$ in our data is $\alpha=3.71\pm0.51$, much lower than for low DM pulsars ($\alpha=3.98\pm0.39$). The small discrepancy between our interpretation and the one given by L04 is clearly a result of the differences between the two datasets. Our database contains much more low-DM pulsars with low $\alpha$ values, and amongst high (or very high) DM pulsars we found significantly more objects with the values of $\alpha$ in the range predicted by simple geometry scattering theory.

The deviations in the values of $\alpha$ are expected for some of the individual sources, as we widely discuss in Paper~1. They are likely due to some unusual scattering geometries (for example in the case of multiple scattering screens or truncated screens, see \citealt{cordes01}). It should be no surprise that such occurrences are less likely for nearby and mid-distance pulsars (hence they do not affect the average value by much), but it would be worth investigating, if they are indeed very common for distant objects, and not just the handful of pulsars we were able to study so far. Our current sample includes only three sources in the DM$>$700~pc~cm$^{-3}$ range, and five pulsars in the 500 to 700~pc~cm$^{-3}$. This statistics clearly asks for an improvement, which would require an observational effort.  New pulse broadening measurements would be extremely useful at frequencies between 1 and 2 GHz, where the scattering for high DM pulsars should be large enough to make pulse broadening measurements reliable. So far we have only one source in this range with exceptionally good coverage - PSR~B1758$-$23 with estimates at nine frequencies between 1.0 and 2.3~GHz (see Paper~2).

Clearly, our results also support the notion that for individual sources the scattering is highly line-of-sight dependant. With this, the understanding of the character of scattering medium in our Galaxy, both in terms of its geometry, as well as its physical (turbulent) properties, will require further investigation and better statistics than 64 lines-of-sight that currently have been studied. Hence the need for further scattering measurements, both for nearby sources - where the new telescope projects such as LOFAR, MWA or LWA may be very helpful - as well as for the distant objects (with very large DM's), for which  measurements with good frequency coverage around 1~GHz would be extremely useful.

\section*{Acknowledgments}
We thank the staff of the GMRT that made these observations possible. GMRT is run by the National Centre for Radio Astrophysics of the Tata Institute of Fundamental Research. We thank the anonymous referee for the comments that helped to significantly improve our paper. WL, KR and JK acknowledge the support of the grant DEC-2013/09/B/ST9/02177 of the Polish National Science Centre. 
This work was also partially supported by grant DEC-2012/05/B/ST9/03924 of the Polish National Science Centre. 
BB acknowledges support of   Marie Curie Actions International Incoming  Fellowship grant no. PIIF-GA-2013-626533. Additional thanks for Rahul Basu, Magdalena Kowali\'nska and Marta Dembska for technical help.

\appendix

\section{Multi frequency pulsar profiles}

Figure~\ref{app_profiles} shows the profiles we collected in our observing project (together with Figure~\ref{sample_profile} this represents all the collected profiles). The red lines on the plots represent the results of modelling used to obtain the pulse broadening times. The apparently lower signal-to-noise at the lowest frequencies - 150 and 235 MHz - is due to both the increased sky background temperature (all of these pulsars are very close to the Galactic plane) as well as to the fact that the observational bandwidth was only half of the value used on higher frequencies. In case of PSR~J1852$-$0635 the low signal-to-noise is also induced by the fact that this object was recently (but after we included it in our sample) identified as a gigahertz-peaked spectrum pulsar \citep[see][]{dembska14}. 

\section{The pulse broadening frequency scaling database}

Table \ref{tab3} presents the results of the scatter time frequency scaling index $\alpha$ from Paper~1 and Paper~2 for which we corrected the uncertainty analysis, using the approach presented in this paper (see Section 3) - i.e. using both the $\chi^2$ mapping technique and the Monte-Carlo simulations. For the majority of these pulsars this did not affect the actual value of $\alpha$ as only the uncertainty of the fit to the $\tau_d$ vs frequency fit was re-calculated. The Table also contains the relevant basic pulsar data (taken from the ATNF Catalogue) , the values of the standardized scattering at 1~GHz and the scattering fluctuation strength $\log C_{n_e}^2 $ - these two parameters were also re-calculated using the method described in this paper.

The data shown in this table, along with the new estimates from Table~\ref{table_alpha} and the estimates of $\alpha$ published elsewhere (for pulsars for which we did not add any new scatter time measurements; see L01, L04, \citealt{johnston98}, \citealt{cordes85}, \citealt{kuzmin02}, \citealt{lewan11}, \citealt{daszuta13}) constitute the entire scatter time frequency scaling index database we are using in our analysis described in Section~3.

\setcounter{section}{1}

\begin{figure}
\resizebox{7.8cm}{!}{\includegraphics{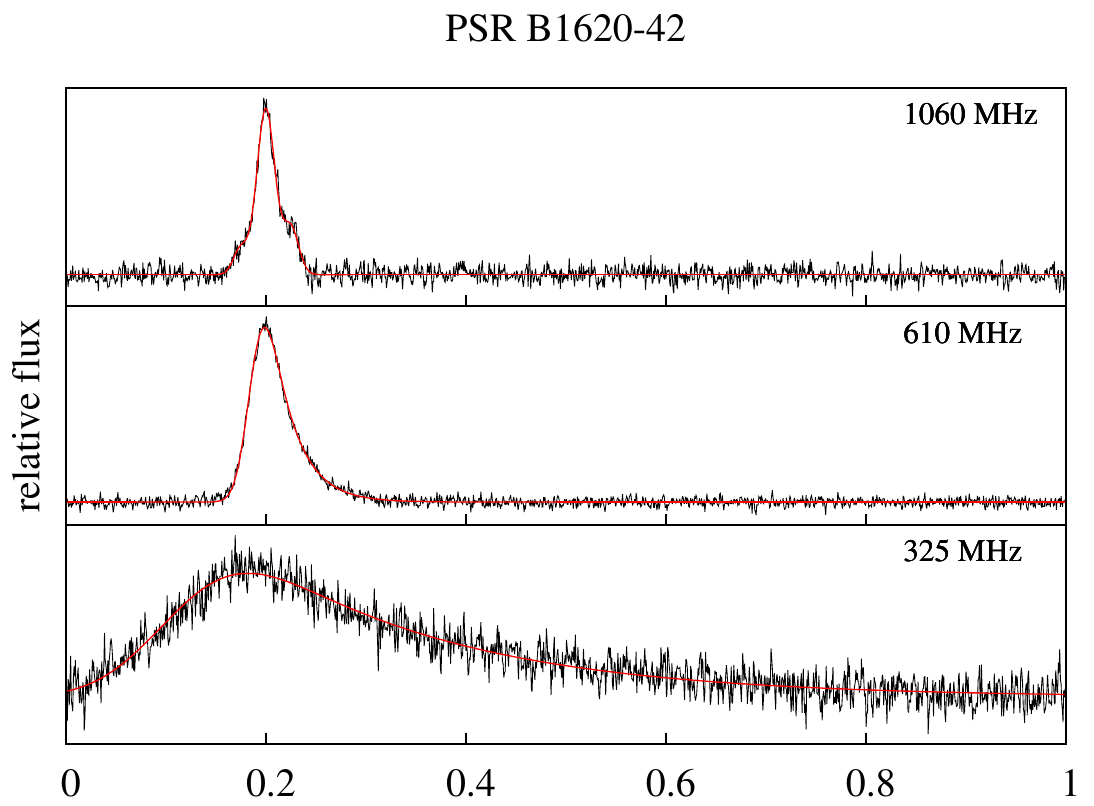}}
\resizebox{7.8cm}{!}{\includegraphics{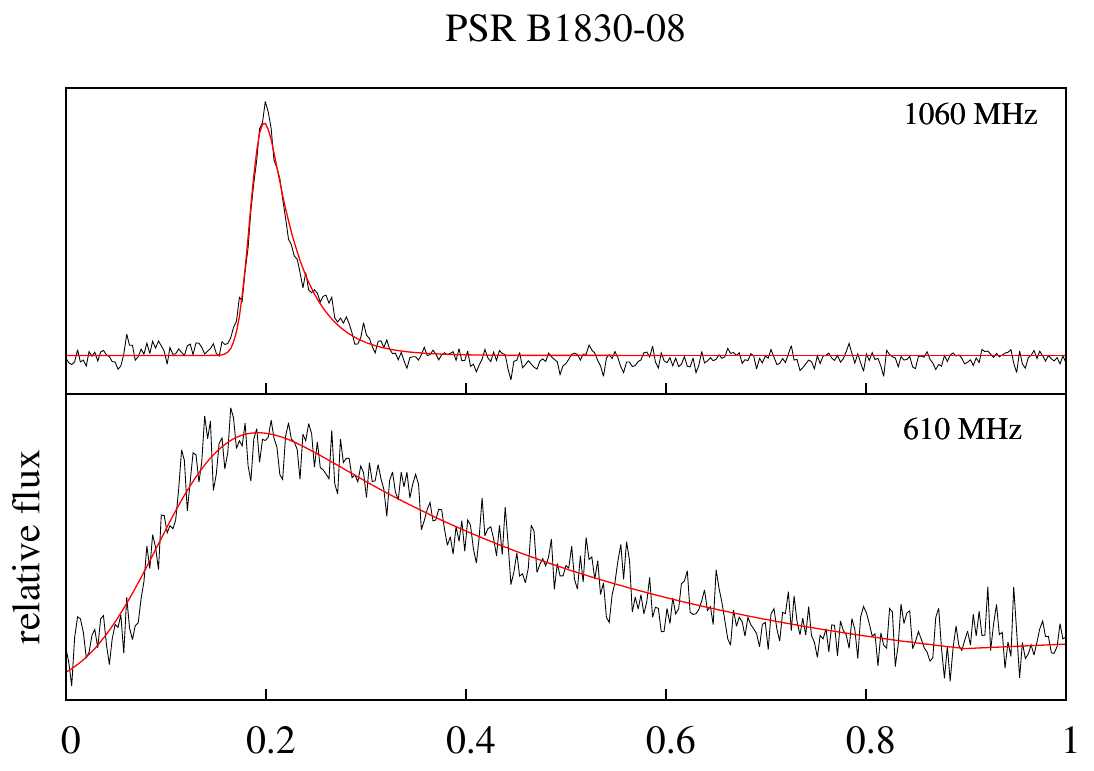}}
\resizebox{7.8cm}{!}{\includegraphics[angle=0]{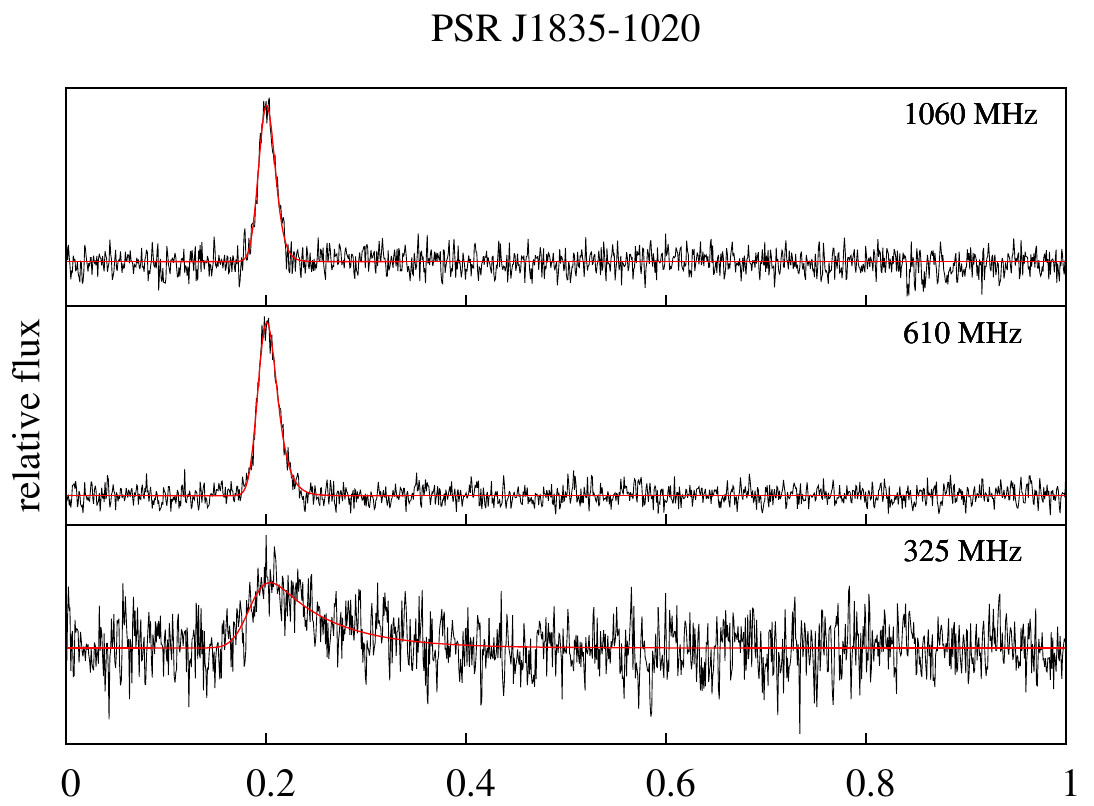}}
\resizebox{7.8cm}{!}{\includegraphics[angle=0]{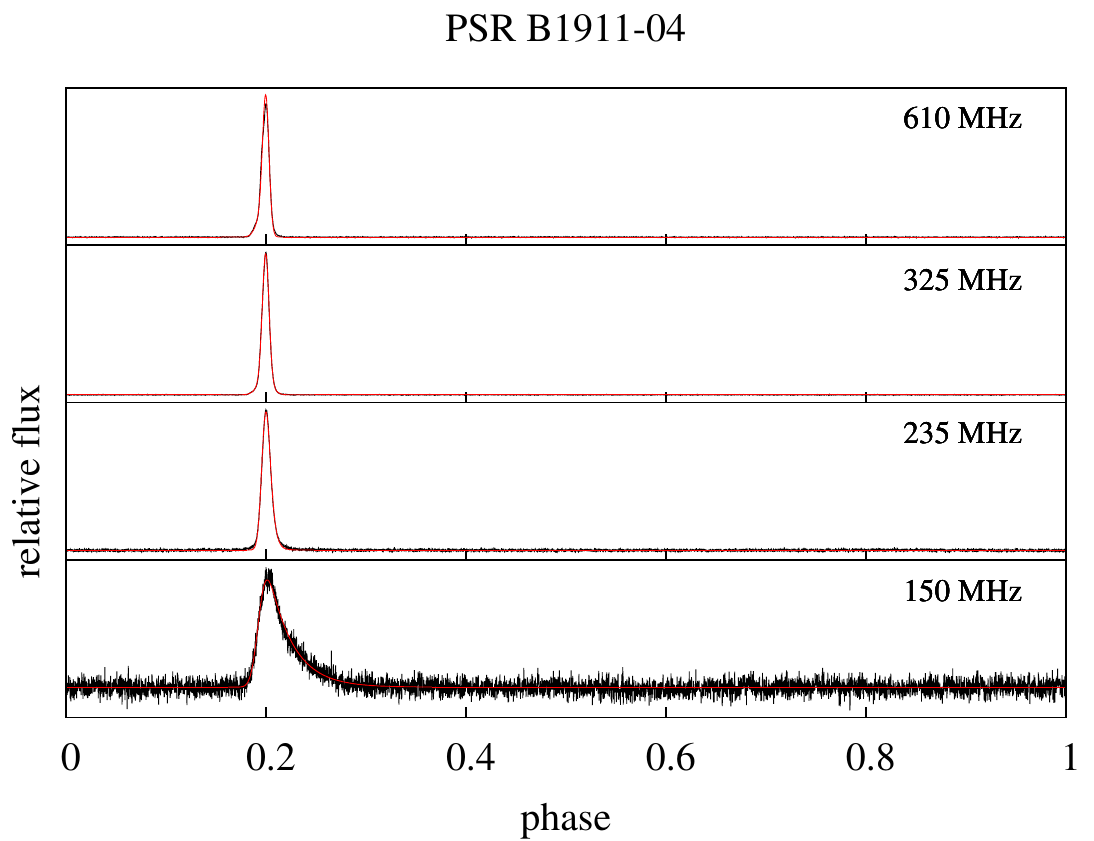}}
\caption{Profiles of pulsars included in our project obtained with GMRT.\label{app_profiles}}
\end{figure}
\setcounter{figure}{0}

\begin{figure}
\resizebox{7.8cm}{!}{\includegraphics[angle=0]{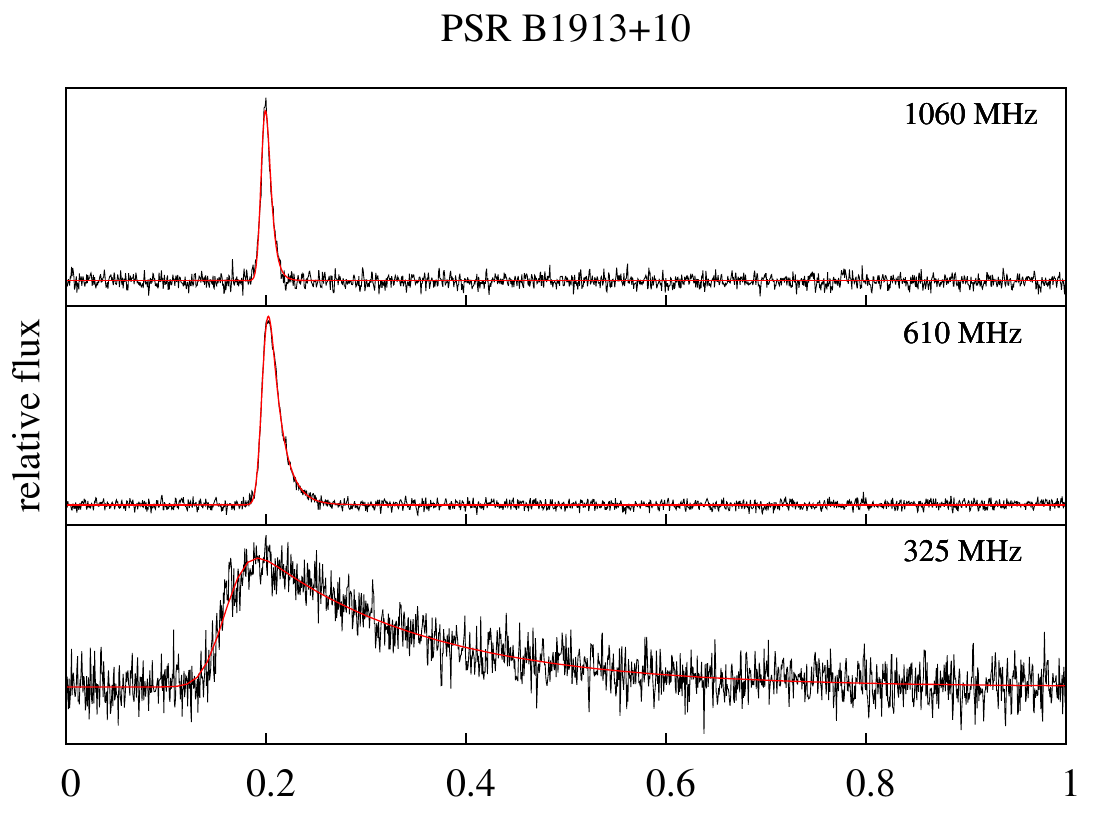}}
\resizebox{7.8cm}{!}{\includegraphics[angle=0]{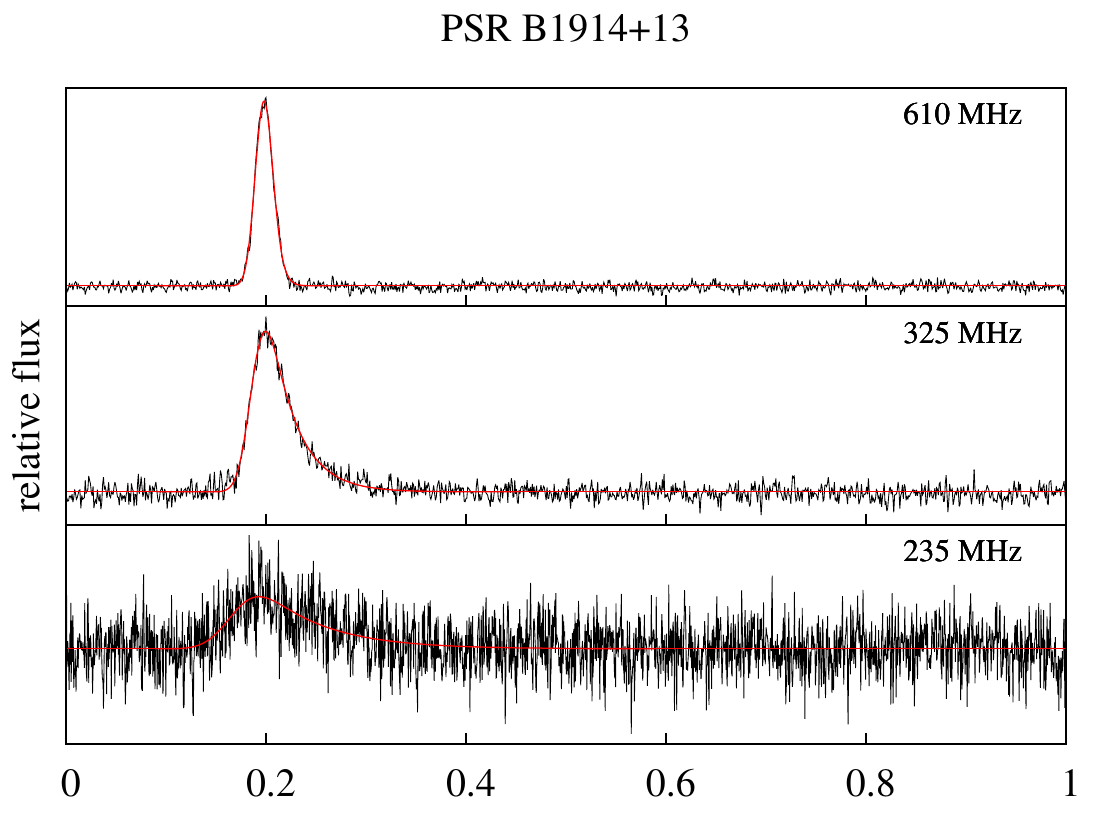}}
\resizebox{7.8cm}{!}{\includegraphics[angle=0]{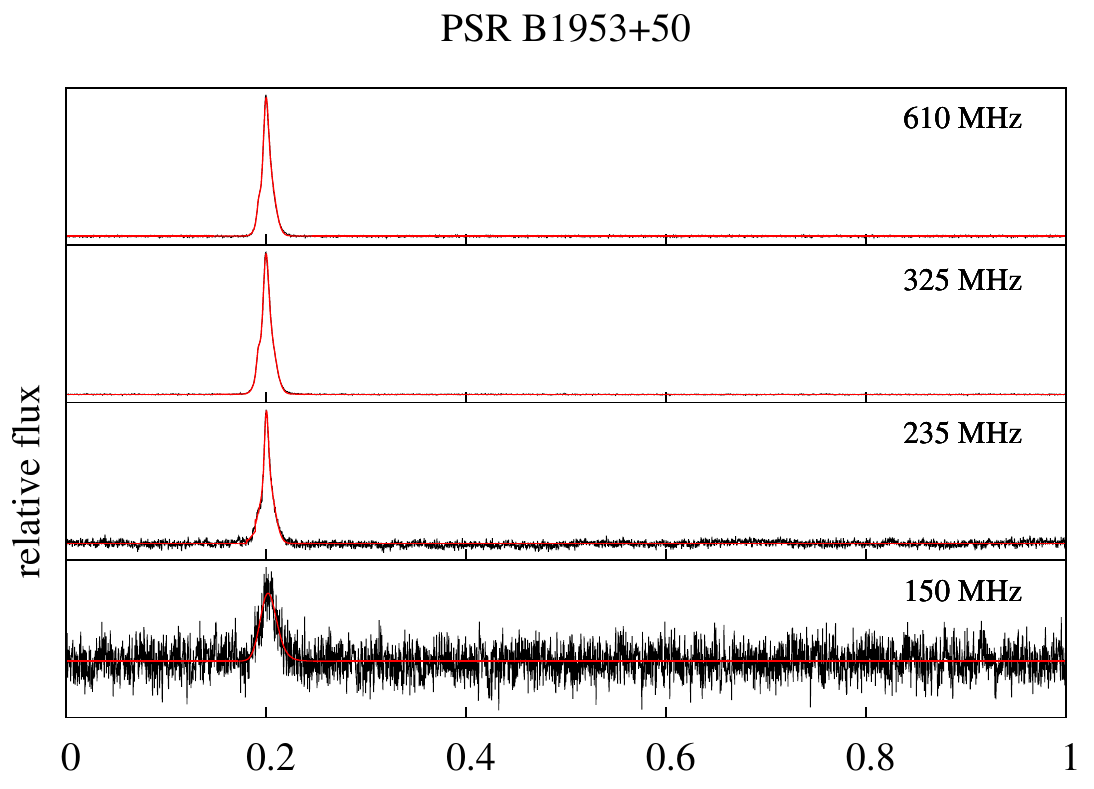}}
\resizebox{7.8cm}{!}{\includegraphics[angle=0]{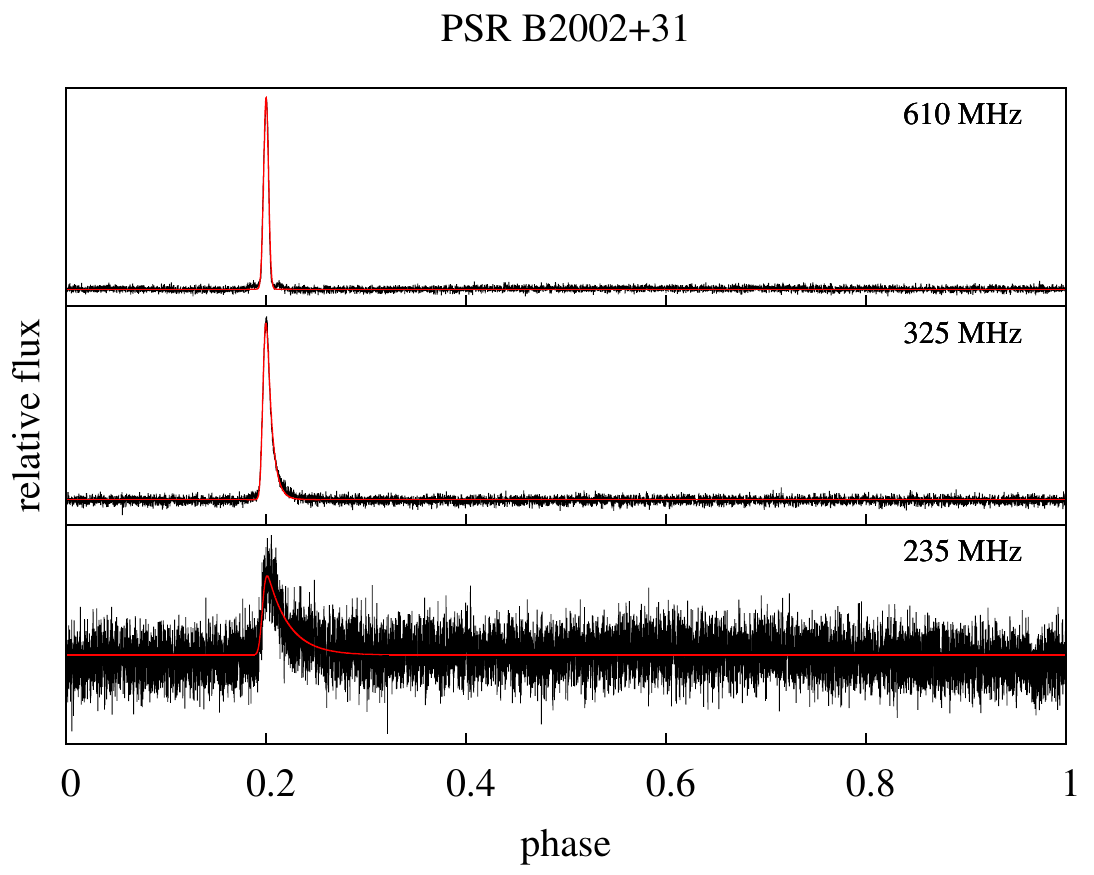}}
\caption{ (continued) Pulsar profiles}
\end{figure}

\clearpage
\setcounter{section}{2}
\begin{table*}
\caption{The scattering spectral index ($\alpha$) and the electron density spectral index ($\beta$) for pulsars from Paper~1 and Paper~2 with corrected uncertainty values (see Section~3 for explanation). Values of $\beta$ indicated by an asterisk are ``un-physical'', as the $\beta=2\alpha/(\alpha-2)$ relation is valid only for $\alpha \geq 4$. Table also lists the scattering fluctuation strength  $\log C_{n_e}^2 $ and the estimated scattering at a standard frequency of 1~GHz.  \label{tab3}}
\def\arraystretch{1.1}
\begin{tabular}{lrrrrlllr}\hline
Pulsar & $l_{II}~~$ &  $b_{II}~~$ & $DM$~~~~~ & Distance & $\alpha$~~ & $\beta$~~  & $\tau_d$ (ms)& $\log C_{n_e}^2 $\\
      & (deg) & (deg) & (pc cm$^{-3}$) & (kpc)~~~ & & & at 1 GHz&\\ \hline \hline
      \multicolumn{9}{c}{Results from Lewandowski et al. (2015)} \\ \hline
    B0339+53&    147.02  &   -1.43   &      67.30   &   2.48  &    3.31$^{+0.24}_{-0.22}$   &     *      &0.0174& $-$3.01\vspace*{0.5mm}\\
    B0402+61&    144.02  &    7.05   &      65.303  &  3.05  &    5.61$^{+0.70}_{-0.58}$\phantom{i}$^{MC}$ & 3.1$^{+0.8}_{-0.7}$      &0.0725& $-$2.86\vspace*{0.5mm}\\
    B0531+21&    184.56  &   -5.78   &      56.791  &  2.00  &    3.67$\pm$0.26$^{MC}$  &     *   &0.00071& $-$4.12\vspace*{0.5mm}\\
    B0808$-$47&    263.30  &   -7.96   &     228.3    &  12.71  &    3.17$\pm$0.22  &     *     &1.94& $-$1.43\vspace*{0.5mm}\\
    B0809+74&    140.00  &   31.62   &       6.116  &   0.43  &    3.9$\pm$1.3 &   4.1$\pm$2.7     & 0.00025& $-$4.76\vspace*{0.5mm}\\
    B0833$-$45&    263.55  &   -2.79   &      67.99   &  0.28  &    4.45$^{+0.76}_{-0.69}$   &    3.63$^{+1.3}_{-1.1}$     &0.047& $-$0.98\vspace*{0.5mm}\\ 
    B0839$-$53&    270.77  &   -7.14   &     156.5    &  7.77  &    4.61$\pm$0.50$^{MC}$  &     3.5$\pm$0.77      &0.304& $-$3.00\vspace*{0.5mm}\\
   B1114$-$41&   284.45   &  18.07    &     40.53    &  2.68   &   3.50$\pm$0.32$^{MC}$   &   *       &0.210& $-$1.66\vspace*{0.5mm}\\
   B1154$-$62&   296.71   &  -0.20    &    325.2     &  4.00   &   4.59$^{+0.33}_{-0.36}$   &    3.54$^{+0.51}_{-0.54}$      &0.493&$-$2.30\vspace*{0.5mm} \\
   B1323$-$58&   307.50   &   3.56    &    287.30    &  3.00   &   4.00$\pm$0.22$^{MC}$   &    4.00$\pm$0.43      &9.47& $-$0.48\vspace*{0.5mm}\\
   B1323$-$62&   307.07   &   0.20    &    318.80    &  4.00   &   5.57$\pm$0.42   &    3.12$\pm$0 .47       &2.35& $-$2.19\vspace*{0.5mm}\\
   B1356$-$60&   311.24   &   1.13    &    293.71    &  5.00   &   3.77$^{+0.61}_{-0.88}$   &    4.3$^{+1.4}_{-1.8}$     &1.04& $-$1.68\vspace*{0.5mm}\\
   B1557$-$50&   330.69   &   1.63    &    260.56    & 6.90   &   3.78$\pm$0.55   &    4.3$\pm$1.3      &5.45&$-$1.17\vspace*{0.5mm} \\
   B1641$-$45&   339.19   &  -0.19    &    478.8     &  4.50   &   3.84$\pm$0.20$^{MC}$   &    4.17$\pm$0.43    &11.0&$-$0.54\vspace*{0.5mm} \\
   J1723$-$3659&   350.68   &  -0.41    &    254.2     &  4.28   &   3.16$^{+0.48}_{-0.44}$\phantom{i}$^{MC}$   &    *     &3.59& 0.32\vspace*{0.5mm} \\
   B1749$-$28&     1.54   &  -0.96    &     50.372   &0.20   &   3.78$\pm$0.27$^{MC}$   &    4.24 $\pm$0.50       &0.0028&$-$1.60\vspace*{0.5mm}\\
     B1758$-$23      &     6.84   &  -0.07    &   1073.9     &  4.00   &   3.45$\pm$0.19   &    *      & 306.9& 2.33\vspace*{0.5mm}\\
   B1821$-$19      &    12.28   &  -3.11    &    224.648   &  3.70   &   3.67$\pm$0.19   &    *      &0.707& $-$1.49\vspace*{0.5mm}\\            
   B1844$-$04      &    28.88   &  -0.94    &    141.979   &  3.12   &   4.71$\pm$0.18$^{MC}$   &    3.48$^{+0.27}_{-0.26}$      & 0.224 &$-$1.80\vspace*{0.5mm}\\
   B1859+03      &    37.21   &  -0.64    &    402.080   &  7.00   &   4.42$\pm$0.35   &    3.65$\pm$0.57      &1.41&$-$2.30\vspace*{0.5mm} \\
   B1900+01      &    35.73   &  -1.96    &    245.167   &  3.30   &   3.53$\pm$0.17   &   *    & 0.58&$-$1.32\vspace*{0.5mm} \\
   B1907+02      &    37.60   &  -2.71    &    171.734   & 4.50   &   3.58$\pm$0.74   &    *     & 0.049& $-$3.05\vspace*{0.5mm}\\
   B1907+10      &    44.83   &   0.99    &    149.982   &  4.80   &   3.61$^{+0.79}_{-0.74}$\phantom{i}$^{MC}$   &    4.4$\pm$2.0    &0.036&$-3.28$\vspace*{0.5mm} \\
   B1919+21      &    55.78   &   3.50    &     12.455   &  0.30   &   3.5$^{+1.0}_{-0.8}$\phantom{i}$^{MC}$   &    4.7$^{+2.6}_{-2.1}$      &0.00031& $-$3.26\vspace*{0.5mm}\\    
   B1933+16      &    52.44   &  -2.09    &    158.521   &  3.70   &   3.35$^{+0.36}_{-0.41}$   &    *     &0.0417& $-$2.89\vspace*{0.5mm}\\
   B1946+35      &    70.70   &   5.05    &    129.075   & 7.87   &   3.63$\pm$0.11$^{MC}$   &  *     &0.619& $-$2.24\vspace*{0.5mm}\\
   B2053+36      &    79.13   &  -5.59    &     97.3140  &  5.00   &   3.78$\pm$0.24$^{MC}$   &   4.14$\pm$0.54      &0.0964& $-2.85$\vspace*{0.5mm}\\
   B2217+47      &    98.38   &  -7.60    &     43.519   &  2.45   &   3.22$^{+0.43}_{-0.50}$\phantom{i}$^{MC}$   &    *     &2.67& $-4.33$\vspace*{0.5mm}\\
   B2303+30      &    97.72   & -26.66    &     49.544   & 3.92   &   3.42$^{+0.47}_{-0.59}$   &    *       &8.50& $-3.94$\vspace*{0.5mm}\\
  \hline
        \multicolumn{9}{c}{Results from Lewandowski et al. (2013)} \\ \hline
     B1740$-$31  &  357.30  &   -1.15     &193.05       &  3.65   &$4.48 ^{+0.31}_{- 0.41}$   & 3.61$^{+0.63}_{-0.83}$       & 1.67&-1.75\vspace*{0.5mm}\\ 
     B1750$-$24  &  4.27 & 0.51             &  672 &  10.18           &4.06$\pm$0.77  & 3.9$\pm$1.5 & 338.5 & -0.10\vspace*{0.5mm} \\
     B1815$-$14  & 16.41 & 0.61            & 622  &   8.10            & 3.97$\pm0.49^{MC}$ & 4.0$\pm0.9$ & 60.9 & -0.48 \vspace*{0.5mm}\\
     B1820$-$14  & 17.25 & -0.18           & 651 &  7.77              & $3.96^{+0.56}_{-0.51}$\phantom{i}$^{MC}$ & 4.0$^{+1.1}_{-1.0}$ & 7.90 & -1.33\vspace*{0.5mm}\\
     B1822$-$14 & 16.81 & -1.00            &357 &  5.45                & $3.77^{+0.54}_{-0.57}$\phantom{i}$^{MC}$ & 4.3$^{+1.2}_{-1.3}$ & 20.22 & -0.31\vspace*{0.5mm} \\
     B1828$-$11 & 20.81 & -0.48 &         161.50 & 3.58            & $3.17^{+0.56}_{-0.34}$\phantom{i}$^{MC}$ & * & 0.98 & -0.45\vspace*{0.5mm} \\
B1832$-$06      & 25.09 & 0.55        &   472.9   &  6.44          & $4.37^{+0.38}_{-0.34}$\phantom{i}$^{MC}$ & 3.69$^{+0.64}_{-0.57}$ & 116.61 & -0.59\vspace*{0.5mm} \\
B1838$-$04 & 7.82   &   0.28 &   325.487 &  5.17                 & 3.91$\pm$0.70$^{MC}$ & 4.1$\pm$1.5 & 3.23 & -1.32\vspace*{0.5mm} \\
J1907+0918 & 43.02    &   0.73 &  357.9  &  7.68                  & $3.6^{+0.9}_{-1.0}$ & 4.4$^{+2.1}_{-2.5}$ & 0.58 & -2.26\vspace*{0.5mm} \\
\hline
\end{tabular}
\end{table*}

\clearpage


\begin{thebibliography}{99}
\bibitem[\protect\citeauthoryear{Bhat et al.}{2003}]{bhat03}
Bhat, N. D. R., Cordes, J. M. \&  Chatterjee, S. 2003, ApJ, 584, 782
\bibitem[\protect\citeauthoryear{Bhat et al.}{2004}]{bhat04}
Bhat, N.D.R., Cordes, J.M., Camilo, F., Nice, D.J. \& Lorimer, D.R. 2004, ApJ, 605, 759
\bibitem[\protect\citeauthoryear{Brisken et al.}{2010}]{brisken10}
Brisken, W.F., Macquart, J.-P., Gao, J.J., Rickett, B.J., Coles, W.A., Deller, A.T., Tingay, S.J. \& West, C.J. 2010, ApJ, 708, 232
\bibitem[\protect\citeauthoryear{Cordes, Weisberg \& Boriakoff}{1985}]{cordes85}
Cordes, J.M., Weisberg, J.M. \& Boriakoff, V. 1985, ApJ, 288, 221
\bibitem[\protect\citeauthoryear{Cordes}{1986}]{cordes86}
Cordes, J.M. 1986, ApJ, 311, 183
\bibitem[\protect\citeauthoryear{Cordes \& Lazio}{2001}]{cordes01}
Cordes, J. M. \& Lazio, T. J. W. 2001, ApJ, 549, 997
\bibitem[\protect\citeauthoryear{Cordes \& Lazio}{2002}]{cordes2002}
Cordes, J. M. \& Lazio, T. J. W. 2002, arXiv:astro-ph/0207156
\bibitem[\protect\citeauthoryear{Daszuta, Lewandowski \& Kijak}{2013}]{daszuta13}
Daszuta, M., Lewandowski, W. \& Kijak, J., 2013, MNRAS, 436, 2492
\bibitem[\protect\citeauthoryear{Dembska~et~al.}{2014}]{dembska14}
Dembska, M., Kijak, J., Jessner, A., Lewandowski, W., Bhattacharyya B. \& Gupta, Y. 2014, MNRAS, 445, 3105
\bibitem[\protect\citeauthoryear{Demorest}{2011}]{demorest11}
Demorest, P.B. 2011, MNRAS, 416, 2821
\bibitem[\protect\citeauthoryear{Johnston, Nicastro \& Koribalski}{1998}]{johnston98}
Johnston, S., Nicastro, L. \& Koribalski, B. 1998, MNRAS, 297, 108
\bibitem[\protect\citeauthoryear{Kijak \& Gil}{2003}]{kijak03}
Kijak, J. \& Gil, J. 2003, A\&A, 397,969
\bibitem[\protect\citeauthoryear{Kijak~et~al.}{2011}]{kijak11}
Kijak, J., Lewandowski, W., Maron, O., Gupta, Y. \& Jessner, A. 2011, A\&A, 531, A16
\bibitem[\protect\citeauthoryear{Krishnakumar~et~al.}{2015}]{krishna15}
Krishnakumar, M.A., Mitra, D., Naidu, A., Joshi, B.C. \& Manoharan P.K. 2015,  ApJ, 804, 9
\bibitem[\protect\citeauthoryear{Kuzmin~et~al.}{2002}]{kuzmin02}
Kuzmin, A.D., Kondratev, V.I, Kostyuk, S.V, Losovsky, B. Ya., Popov, M. V., Soglasnov, V. A., D'Amico, N., \& Montebugnoli, S. 2002, Astron. Lett. 28, 251
\bibitem[\protect\citeauthoryear{Lambert \& Rickett}{1999}]{lambert99}
Lambert, H.C. \& Rickett, B.J. 1999, ApJ, 517, 299
\bibitem[\protect\citeauthoryear{Lewandowski~et~al.}{2011}]{lewan11}
Lewandowski, W., Kijak, J., Gupta, Y. \& Krzeszowski, K. 2011, A\&A, 534, A66
\bibitem[\protect\citeauthoryear{Lewandowski~et~al.}{2013}]{lewan13}
Lewandowski, W., Dembska, M., Kijak, J. and Kowali\'nska, M. 2013, MNRAS, 434, 69 {\bf (Paper~1)}
\bibitem[\protect\citeauthoryear{Lewandowski~et~al.}{2015}]{lewan15}
Lewandowski, Kowali\'nska, M. \& Kijak, J.  2015, MNRAS, 449, 1570 {\bf (Paper~2)}
\bibitem[\protect\citeauthoryear{L{\"o}hmer~et~al.}{2001}]{L01}
L{\"o}hmer, O., Kramer, M., Mitra, D., Lorimer, D.R. \& Lyne, A.G. 2001, ApJ, 562, L157 {\bf (L01)}
\bibitem[\protect\citeauthoryear{L{\"o}hmer~et~al.}{2004}]{L04}
L{\"o}hmer, O., Mitra, D., Gupta, Y., Kramer, M. \& Ahuja, A. 2004, A\&A, 425, 569 {\bf (L04)}
\bibitem[\protect\citeauthoryear{Manchester~et~al.}{2005}]{manchester05}
Manchester, R. N., Hobbs, G. B., Teoh, A. \& Hobbs, M. 2005, AJ, 129, 1993-2006
\bibitem[\protect\citeauthoryear{Ramachandran~et~al.}{1997}]{rama97}
Ramachandran, R., Mitra, D., Deshpande, A. A., McConnell, D. M. \& 
Ables, J. G. 1997, MNRAS, 290, 260
\bibitem[\protect\citeauthoryear{Rickett}{1977}]{rickett77}
Rickett, B.J. 1977, ARA\&A, 15, 479
\bibitem[\protect\citeauthoryear{Rickett}{1990}]{rickett90}
Rickett, B.J. 1990, ARA\&A, 28, 561
\bibitem[\protect\citeauthoryear{Rickett~et~al.}{2009}]{rickett09}
Rickett, B., Johnston, S., Tomlinson, T. \& Reynolds, J. 2009, MNRAS, 395, 1391
\bibitem[\protect\citeauthoryear{Romani, Narayan \& Blandford}{1986}]{romani86}
Romani, R.W., Narayan, R. \& Blandford, R. 1986, 220, 19
\bibitem[\protect\citeauthoryear{Roy et al.}{2010}]{roy2010}
Roy, J., Gupta, Y., Ue-Li Pen et al, 2010, Experimental Astronomy, 28, 25
\bibitem[\protect\citeauthoryear{Scheuer}{1968}]{scheuer68}
Scheuer, P.A.G. 1968, Nature, 218, 920
\bibitem[\protect\citeauthoryear{Williamson}{1972}]{williamson72}
Williamson, I.P. 1972, MNRAS, 157,55
\bibitem[\protect\citeauthoryear{Williamson}{1973}]{williamson73}
Williamson, I.P. 1973, MNRAS, 163,345
\end{thebibliography}
\end{document}